\documentclass[manuscript,screen,nonacm]{acmart}
\renewcommand\footnotetextcopyrightpermission[1]{} 
\AtBeginDocument{%
  }

\setcopyright{none}
\copyrightyear{2025}
\acmDOI{XXXXXXX.XXXXXXX}

\usepackage{multirow}
\usepackage{spverbatim}
\usepackage{algorithmic}
\usepackage{algorithm}
\usepackage{amsmath}
\begin{document}

\title{The Personality Trap: How LLMs Embed Bias When Generating Human-Like Personas}


\author{Jacopo Amidei}
\affiliation{%
  \institution{Universitat Oberta de Catalunya}
  \city{Barcelona}
  \country{Spain}}
\email{jamidei@uoc.edu}

\author{Gregorio Ferreira}
\affiliation{%
  \institution{Universitat Oberta de Catalunya}
  \city{Barcelona}
  \country{Spain}}
\email{jferreirade@uoc.edu}

\author{Mario Mu\~{n}oz Serrano}
\affiliation{%
  \institution{Universitat Oberta de Catalunya}
  \city{Barcelona}
  \country{Spain}}
\email{mariomunozserrano@gmail.com}

\author{Rub\'{e}n Nieto}
\affiliation{%
  \institution{Universitat Oberta de Catalunya}
  \city{Barcelona}
  \country{Spain}}
\email{rnietol@uoc.edu}

\author{Andreas Kaltenbrunner}
\affiliation{%
  \institution{Universitat Pompeu Fabra}
  \city{Barcelona}
  \country{Spain}}
\email{andreas.kaltenbrunner@upf.edu}

\begin{abstract}

This paper examines biases in large language models (LLMs) when generating synthetic populations from responses to personality questionnaires. 
Using five LLMs, we first assess the representativeness and potential biases in the sociodemographic attributes of the generated personas, as well as their alignment with the intended personality traits. While LLMs successfully reproduce known correlations between personality and sociodemographic variables, all models exhibit pronounced WEIRD (western, educated, industrialized, rich and democratic) biases—favoring young, educated, white, heterosexual, Western individuals with centrist or progressive political views and secular or Christian beliefs. In a second analysis, we manipulate input traits to maximize Neuroticism and Psychoticism scores. Notably, when Psychoticism is maximized, several models produce an overrepresentation of non-binary and LGBTQ+ identities, raising concerns about stereotyping and the potential pathologization of marginalized groups. Our findings highlight both the potential and the risks of using LLMs to generate psychologically grounded synthetic populations.
\end{abstract}





\maketitle
\fancyfoot{}
\pagestyle{plain}

\section{Introduction}
The use of large language models (LLMs) to generate synthetic populations 
offers a powerful, low-cost method for simulating human behavior.
The ability to generate synthetic personas is being increasingly explored in fields ranging from psychology and political science to software engineering and healthcare. Synthetic personas,
referred to by various names such as \textit{guinea pigbots} \cite{hutson2023guinea}, \textit{silicon samples}, or \textit{homo silicus} \cite{horton2023large}, 
have been proposed as substitutes for human participants in experimental research and as stand-ins for survey respondents or user testers.
However, despite their growing popularity, important questions remain about how representative and unbiased these synthetic populations actually are.
Recent evidence suggests that these representativeness concerns are not only a property of the underlying model, but also of how personas are elicited: persona prompting strategies can substantially change portrayals and stereotyping, and LLMs often struggle more when simulating marginalized groups \cite{lutz-etal-2025-prompt}. 
Moreover, bias can be amplified when personas combine multiple traits in ways that are unlikely in real survey data (``incongruous personas''), leading to systematic deviations in persona-steered generation \cite{liu-etal-2024-evaluating-large}.

Synthetic populations are usually defined based on demographic  \cite{argyle2023out, ferreira2024matching, sun2024random}, or individual-level input data \cite{park2022social} or giving information about historical or imaginary well-known persons \cite{shao2023character},
but here we investigate a different and under-explored method: generating populations based solely on personality test scores.
The rationale for using personality traits stems from their well-established correlations with a wide range of social and behavioural outcomes, from
moral judgments \cite{sun2024moral, luke2022big, andrejevic2022basic, schwartz2021association} as well as value systems \cite{czerniawska2021values}, civic engagement \cite{stahlmann2024big}, preferences for and voting for green parties \cite{bleidorn2024high}, vaccination attitudes, intentions, and behaviours \cite{bleidorn2024big} 
to ethical vegetarianism and attitudes about animal welfare legislation \cite{smillie2024differential, trenkenschuh2024personality}. 

Leveraging these relationships, we examine how well LLMs can generate synthetic populations that reflect both the intended personality profiles and associated (if possible, unbiased) sociodemographic distributions. We pose three key research questions:

\begin{itemize}
    \item[RQ1:] Which sociodemographic attributes do LLMs generate when personas are based solely on personality test responses, and to what extent do these outputs reflect demographic biases?
    \item[RQ2:] How does trait manipulation (e.g., maximizing Neuroticism or Psychoticism) influence demographic outputs?
    \item[RQ3:] To what extent do LLM-generated personas internalize and express input personality traits? 
\end{itemize}
To address these questions, we use responses from the Eysenck Personality Questionnaire Revised-Abbreviated (EPQR-A)~\cite{Francis1992} as inputs to five state-of-the-art LLMs and generate synthetic population samples. 

To answer RQ1, we prompt the LLMs to provide sociodemographic attributes (e.g., gender, occupation, and political orientation) when generating the synthetic personas. Potential biases in the representativeness of the sample population can then be assessed by examining the distributions of these attributes. The same strategy was applied for RQ2, where the effect of extreme trait manipulation on the resulting generated personas is studied.  Finally, to address RQ3, we employed multiple strategies: (1) measure the correlation between two comparable personality tests based on each synthetic persona’s responses; (2) assess accuracy metrics by comparing synthetic responses to those of the input population; and (3) measure the internal consistency of the personality test employed.


Our findings suggest that LLMs are capable of generating synthetic sample populations that mirror to a large extent the input personality traits. However, the results also consistently reveal significant WEIRD biases. Synthetic populations skew heavily toward young, educated, white, Western, heterosexual individuals with secular or Christian backgrounds and progressive political views. More concerningly, we find that maximizing traits like Psychoticism leads to a marked overrepresentation of LGBTQ+ and non-binary identities in several models, raising important questions about harmful stereotypes and potential pathologization.

By combining psychometric rigor with generative analysis, this study contributes new insights into both the capabilities and risks of using LLMs to simulate human populations, especially when these models are treated as proxies for real-world diversity.

\section{Related work}

This paper bridges three research areas in the LLM framework: i) Building synthetic sample populations with LLMs, ii) Assessing LLM personality through the use of personality tests, and iii) Bias detection in LLMs.

\subsection{Building synthetic sample populations with LLMs}
The employment of LLMs as substitutes for human participants was studied in psychological research \cite{hutson2023guinea, dillion2023can, park2024diminished},  political polling \cite{sanders2023demonstrations}, software engineering research \cite{gerosa2024can}, teaching research \cite{markel2023gpteach}, economics~\cite{horton2023large}, social media platforms design \cite{park2022social, tornberg2023simulating}, market research to understand consumer preferences \cite{brand2023using} and more generally social science research \cite{argyle2023out}. 
Across these domains, findings are mixed: some studies report alignment with human response patterns \cite{argyle2023out, ma2025algorithmic, park2024generative, sanders2023demonstrations, park2022social, horton2023large, sun2024random}, whereas others caution that LLM substitutes can fail in systematic ways and should not be treated as drop-in replacements for human participants 
\cite{crockett2023should, park2024diminished, agnew2024illusion, wang2024large, harding2023ai, petrov2024limited, ai2024self}.
For example, \citet{agnew2024illusion} sheds light on potential obstacles when utilizing LLMs to simulate human behaviour, such as the current inability of LLMs to emulate human cognition and decision-making accurately, the reliance of psychology research on various non-linguistic cues to study human cognition and behaviour, and the phenomenon of ``value lock-in'' (that is the LLMs ability of reflecting attitudes only from the time of their training). 

Recent work has also begun to test whether synthetic participants generalize beyond WEIRD settings and policy domains \cite{shrestha2024beyond}, and methodological discussions have started to formalize LLMs as artificial research participants and clarify the risks of substituting humans with model-generated respondents in behavioral research designs \cite{medina2025artificialparticipants}.



\subsection{Assessing LLM personality through the use of personality tests}
In recent years, there has been a surge in the employment of personality questionnaires within the LLM framework. For example, the Big Five factors \cite{digman1990personality} were used, among others, by \citet{karra2022estimating, safdari2023personality, pellert2023ai, mei2024turing, jiang2024evaluating} to quantify the personality traits of LLMs. Similarly, IPIP-NEO \cite{goldberg1999broad} was used in \citet{safdari2023personality} and Short Dark Tetrad (SD4) \cite{paulhus2020screening} was used in \citet{pellert2023ai}. The EPQR-A was instead used in a multilingual setting in \citet{amidei2025exploring, dewell, ferreira2024matching}. 
While the outcomes of the aforementioned studies may vary depending on the LLMs and questionnaires used, there is support to conclude that personality assessments for LLMs are valid and reliable. Nevertheless, \citet{dorner2023personality, zou2024can, gupta2024selfassessment} argue against the use of self-reported tests for LLMs.

An important validity concern in questionnaire-based profiling is that LLMs may treat survey administration as an evaluation context and respond in systematically self-presentational ways. 
 \citet{salecha2024large} show that across multiple LLM families, responses to Big Five surveys shift toward socially desirable trait poles once models can infer that they are being evaluated.
This line of work suggests that part of what psychometric instruments measure in LLMs can reflect emergent social-desirability responding rather than stable underlying traits, which is particularly salient for instruments (such as EPQR-A) that include explicit social-desirability components.

\subsection{Bias detection in LLMs}
As LLMs are increasingly used to generate text that stands in for people (e.g., synthetic respondents or personas), auditing demographic bias and representational harms becomes essential. In this work, we focus specifically on bias that emerges in LLM-generated synthetic populations.
For a more comprehensive discussion on bias in LLMs, we refer the reader to \cite{gallegos2024bias, guo2024bias, kumar2025no}.
Concerns about the use of synthetic sample populations, due to inherent biases in LLMs, have been raised by, among others, in \cite{crockett2023should, agnew2024illusion, wang2024large, harding2023ai}. For example, \citet{crockett2023should} points up the problem of population generability. This problem was studied by \citet{harding2023ai}, who criticized the fact that LLMs' population representativeness must be carefully circumscribed. \citet{wang2024large} further show that LLMs misportray and flatten demographic groups due to intrinsic model bias and limitations, including their difficulty in representing
minority groups \cite{agnew2024illusion}. Indeed, the problem of misrepresentation, defined as an incomplete or non-representative distribution of a sample population generalised to a broader social group, is a common form of social bias identified in NLP \cite{gallegos2024bias}. 

Relatedly, recent NLP work has studied bias under persona-steered generation, showing that bias is not only a property of model parameters but also of how personas are composed and prompted. 
\citet{liu-etal-2024-evaluating-large} introduce the notion of incongruous personas (multi-trait personas whose traits are unlikely to co-occur in human data) and show that LLM generations can deviate from expected distributions when personas combine multiple identity-relevant attributes. 
Complementing this, \citet{lutz-etal-2025-prompt} systematically evaluate sociodemographic persona prompting strategies and find that seemingly minor prompt choices (e.g., role-adoption format or demographic priming) can substantially change portrayals and stereotyping, particularly for marginalized groups.
Finally,  \citet{ostrow-lopez-2025-llms} document that LLMs reproduce stereotypes about sexual and gender minorities beyond binary gender categories, both in survey-style elicitation and in free-form text generation, raising concerns about representational harms.



\subsection{Research gap}
To the best of our knowledge, prior work has not examined the generation of synthetic populations conditioned solely on responses (or scores) from personality questionnaires. 
This gap matters because personality-test conditioning is an appealingly lightweight way to generate large-scale personas without collecting biographical data. Yet it may also create a direct channel for stereotype activation when downstream demographic attributes are inferred from the personality-conditioned text. 
We therefore integrate the three strands above to evaluate both (i) whether personality-score conditioning yields coherent and diverse personas, and (ii) whether it introduces systematic demographic distortions and representational harms.

\section{Methods}
\subsection{Experimental Setups}\label{subsec:experimental_setups}

\begin{figure}[t!]
\centering
\includegraphics[width=\textwidth,trim={0 0.2cm 0 0},clip]{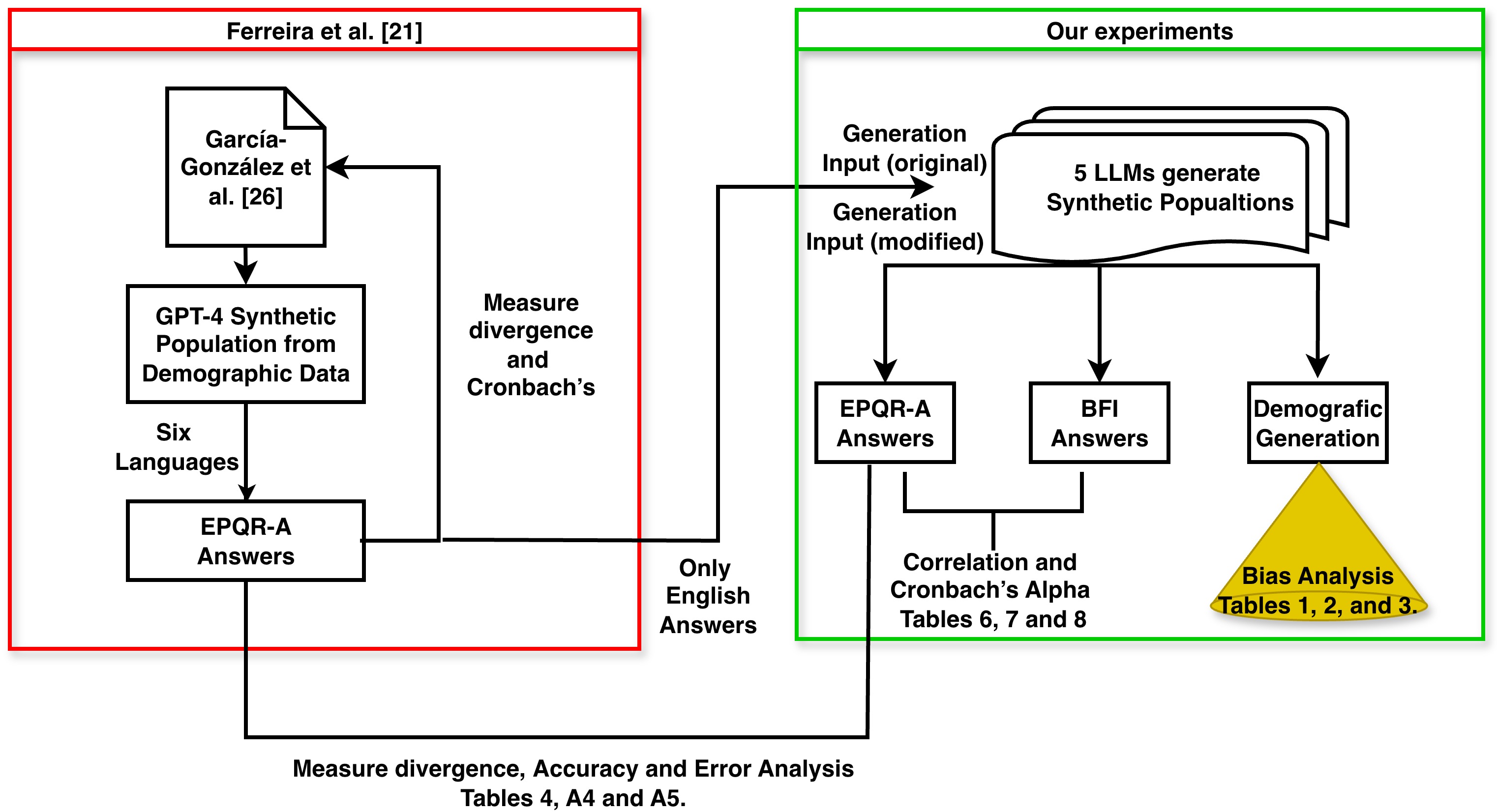}
\caption{Overview of the experimental pipeline. Our experiment (right square) builds on the English version of the EPQR-A responses from 826 simulated personas reported in \citet{ferreira2025well} (left square), which were used as generation input. Five LLMs were then prompted to generate 826 synthetic personas and provide sociodemographic attributes while reflecting the input personality traits. Subsequently, the LLMs were tasked to answer both the EPQR-A and BFI with the generated personalities.
In a further step, the LLMs generated another 826 synthetic personas based on extreme trait manipulations, again producing sociodemographic attributes consistent with the input traits. Potential biases in the representativeness of the sample populations were assessed by analysing the distributions of these attributes. Divergence, accuracy, and error analysis were computed between the input responses and those newly generated. In addition, correlations between the EPQR-A and the BFI were examined, along with internal consistency measured using Cronbach’s alpha.}  
\label{fig:explanatory_figure}
\end{figure}

We designed a multi-step experimental pipeline combining psychometric profiling, generative prompting, and statistical evaluation.\footnote{The code used for our experiments can be found at \url{https://anonymous.4open.science/r/the_personality_trap-F487/README.md}.}  The different steps of our pipeline are visualized in Figure~\ref{fig:explanatory_figure}. 

The input to our study is a dataset of 826 simulated responses to the EPQR-A personality questionnaire, provided by \citet{ferreira2025well}\footnote{The synthetic populations generated in \citet{ferreira2025well} were based on the demographic characteristics (655 females: mean age = 18.9, SD = 1.56 and 171 males: mean age = 19.36, SD = 1.99) of a population described in \citet{garcia2021eysenck}.}. The authors of this study used different questionnaire languages and parameter settings. Here we only use the responses to the English version of the EPQR-A completed by 
gpt-4o-2024-05-13  with temperature 1.
Each of these responses was used as input to prompt an LLM to generate a detailed synthetic persona, which should represent a synthetic persona whose demographic, biographical, and behavioral features were to reflect the personality traits inferred from the questionnaire.


   
    

We tested five large language models: GPT-3.5 (gpt3.5-turbo-0125), GPT-4o (gpt4o-2024-11-20), claude-3.5-s (claude3.5-sonnet), and the two open-source models LLaMa3.2-3B and LLaMa3.1-70B.\footnote{We performed our experiments for the GPT model family with OpenAI's API (total cost of 228.12 USD) and for the remaining models on Amazon AWS (total cost of 223.96 USD).} Each model was prompted with a consistent template that embedded the EPQR-A responses and instructed the model to imagine and describe a person based on that personality profile. The output was required in a structured JSON format containing a predefined set of 8 sociodemographic attributes: age, gender, sexual orientation, race,  religious belief, occupation, political orientation, and location. See  
Appendix \ref{appendix:persona_prompt} 
for the corresponding prompts and 
Appendix \ref{appendix:persona_description} 
for examples of the resulting synthetic persona descriptions.

In addition to this unaltered or ``baseline`` generation condition (to which we refer to as Base population in the remainder of this study), we generated a random control baseline condition and two manipulated conditions (MaxN and MaxP) to assess how models respond to extreme personality trait anchoring. The random condition was generated by computing the true/false binomial distributions for each question and then randomly generating 826 answered questionnaires using these distributions. For the MaxN condition, we adjusted each input response to yield the highest possible Neuroticism score while keeping other traits unchanged; similarly, for the MaxP condition, we maximized Psychoticism.
MaxN and MaxP enabled us to investigate how exaggerated personality traits impact sociodemographic outcomes and whether they elicit stereotypical or pathologising patterns.

\subsection{Strategies for Result Evaluation}
To evaluate generation consistency, we generated 10 Base sample populations and 5  MaxN and MaxP 
sample populations using identical personality measures as inputs (i.e., answers to the EPQR-A questionnaire). To determine whether the generated personas reflected meaningful and coherent demographic patterns (RQ1 and RQ2), we analyzed five key identity categories (Gender, Race, Religion, Political, and Sexual Orientation) using descriptive statistics and two-sided t-tests to compare means between Base and MaxN and MaxP populations.

To assess trait fidelity and alignment (RQ3), we conducted another round of testing in which the generated personas were asked to complete the EPQR-A questionnaire again. Since the Inter-sample variation was minimal, we selected one representative population sample from each condition (Base, MaxN, and MaxP) for this analysis.
We then computed accuracy metrics such as Mean Absolute Error (MAE) and Root Mean Squared Error (RMSE) between the input and regenerated EPQR-A scores. 
For the Base sample, we also administered the Big Five personality inventory (BFI) \cite{john1999big}.
Pearson correlations between EPQR-A and BFI scores were calculated to examine cross-instrument consistency, and Cronbach’s alpha \cite{cronbach1951coefficient} was used to assess internal reliability for both EPQR-A and the BFI.

\subsection{Personality tests used in the experiments}
The EPQR-A is an abbreviated version of the Eysenck Personality Inventory \cite{eysenck1964eysenck}, containing 24 items to assess three personality dimensions (6 items each): {\bf Extraversion (E), Neuroticism (N), and Psychoticism (P)}. It also includes a scale to assess social desirability {\bf Lie (L)}, which also contains six items.  Each item has a dichotomous response (yes or no), and a score for each scale can be computed by summing individual items (resulting in a range from 0 to 6). 

The BFI contains 44 items (assessed in a scale from 1 to 5) that measure individuals in the following general dimensions: {\bf Extraversion (E), Agreeableness (A), Conscientiousness (C), Neuroticism (N)}, and {\bf Openness (O)}. Both the EPQR-A and the BFI assess N and E, and taking into account available results in the literature, we hypothesize that scores from the corresponding dimensions of the two questionnaires should be highly and significantly correlated \cite{aziz2001comparison}. Conversely, the dimensions C and A of the BFI are expected to be inversely correlated with the dimension P from the EPQR-A, while the dimension O from the BFI should be positively correlated with E from the EPQR-A \cite{saggino2000big}.

\section{Results}

\begin{table*}[!t]
\small
\centering
  \caption{Averages and standard deviations of ten Base, five MaxN, and five MaxP sample populations statistics for GPT-4o and GPT-3.5. Non-bin., Con., Prog. Hetero. and Unspe. stands for non-binary, conservative, progressive, heterosexual, and unspecified, respectively. Cases where results are significantly different from the Base sample populations are marked as p<0.05*, p <0.01$\dagger$, and p <0.001$\ddagger$ (two-sided t-test).}
  \label{tab:populations_stats_gpt}
    \begin{tabular}{@{}l@{\hskip 1pt}|l@{\hskip 1pt}||r@{\hskip 1pt}|r@{}c@{\hskip 1pt}|r@{}c@{}||r@{\hskip 1pt}|r@{}c@{\hskip 1pt}|r@{}c@{}}
& 
   &
  \multicolumn{5}{c||}{GPT-4o} &
  \multicolumn{5}{c}{GPT-3.5} \\
  \hline

& 
   &
  \multicolumn{1}{c|}{Base} & 
  \multicolumn{2}{c|}{MaxN} & 
  \multicolumn{2}{c||}{MaxP} &
  \multicolumn{1}{c|}{Base} & 
  \multicolumn{2}{c|}{MaxN} & 
  \multicolumn{2}{c}{MaxP}\\ \hline

\hline
     
\multirow{4}{*}{\rotatebox[origin=c]{90}{Gender}} & 
  Female &
  {25.71 $\pm$ 1.36} &
  29.54 $\pm$ 2.07 &
  $\ddagger$ &
  4.67 $\pm$ 0.79 &
  $\ddagger$ &
  {90.82 $\pm$ 2.13} &
  90.94 $\pm$ 1.21 &
   &
  88.74 $\pm$ 1.02 &
  $\dagger$ \\
 &
  Male &
  {44.75 $\pm$ 2.42} &
  30.46 $\pm$ 2.33 &
  $\ddagger$ &
  6.27 $\pm$ 0.89 &
  $\ddagger$ &
  {9.03 $\pm$ 2.28} &
  8.86 $\pm$ 1.14 &
   &
  10.94 $\pm$ 1.00 &
  $\dagger$ \\
 &
  Non-bin. &
  {29.18 $\pm$ 1.76} &
  39.71 $\pm$ 1.27 &
  $\ddagger$ &
  88.76 $\pm$ 0.99 &
  $\ddagger$ &
  {0.14 $\pm$ 0.21} &
  0.19 $\pm$ 0.14 &
   &
  0.31 $\pm$ 0.18 &
   \\
 &
  Other &
  {0.36 $\pm$ 0.21} &
  0.29 $\pm$ 0.22 &
   &
  0.29 $\pm$ 0.20 &
   &
  {0.00 $\pm$ 0.00} &
  0.00 $\pm$ 0.00 &
   &
  0.00 $\pm$ 0.00 &
   \\ \hline
\multirow{4}{*}{\rotatebox[origin=c]{90}{Pol. Or.}} &
  Centre &
  {64.50 $\pm$ 1.43} &
  51.45 $\pm$ 1.47 &
  $\ddagger$ &
  0.36 $\pm$ 0.17 &
  $\ddagger$ &
  {75.08 $\pm$ 0.89} &
  72.98 $\pm$ 1.68 &
  * &
  18.23 $\pm$ 1.13 &
  $\ddagger$ \\
 &
  Con. &
  {0.12 $\pm$ 0.08} &
  0.02 $\pm$ 0.05 &
   &
  0.00 $\pm$ 0.00 &
  * &
  {0.00 $\pm$ 0.00} &
  0.00 $\pm$ 0.00 &
   &
  0.00 $\pm$ 0.00 &
   \\
 &
  Prog. &
  {34.63 $\pm$ 1.44} &
  48.11 $\pm$ 1.46 &
  $\ddagger$ &
  97.70 $\pm$ 0.19 &
  $\ddagger$ &
  {16.46 $\pm$ 1.79} &
  18.55 $\pm$ 1.56 &
  * &
  38.74 $\pm$ 1.95 &
  $\ddagger$ \\
 &
  Others &
  {0.75 $\pm$ 0.13} &
  0.41 $\pm$ 0.22 &
  * &
  1.94 $\pm$ 0.21 &
  $\ddagger$ &
  {8.45 $\pm$ 1.68} &
  8.48 $\pm$ 0.67 &
   &
  43.03 $\pm$ 1.39 &
  $\ddagger$ \\ \hline
\multirow{5}{*}{\rotatebox[origin=c]{90}{Race}} &
  Asian &
  {0.58 $\pm$ 0.29} &
  0.63 $\pm$ 0.34 &
   &
  1.89 $\pm$ 0.53 &
  $\ddagger$ &
  {3.10 $\pm$ 0.41} &
  2.88 $\pm$ 0.40 &
   &
  2.30 $\pm$ 0.24 &
  * \\
 &
  Black &
  {0.05 $\pm$ 0.07} &
  0.00 $\pm$ 0.00 &
   &
  0.29 $\pm$ 0.18 &
  $\dagger$ &
  {0.97 $\pm$ 0.33} &
  0.87 $\pm$ 0.16 &
   &
  1.16 $\pm$ 0.55 &
   \\
 &
  Latin &
  {0.12 $\pm$ 0.00} &
  0.19 $\pm$ 0.11 &
   &
  0.46 $\pm$ 0.23 &
  $\dagger$ &
  {7.33 $\pm$ 1.64} &
  8.26 $\pm$ 0.96 &
   &
  6.10 $\pm$ 0.66 &
  * \\
 &
  White &
  {98.98 $\pm$ 0.43} &
  98.64 $\pm$ 0.43 &
   &
  90.39 $\pm$ 1.00 &
  $\ddagger$ &
  {88.59 $\pm$ 1.83} &
  87.94 $\pm$ 0.83 &
   &
  90.41 $\pm$ 0.58 &
  $\dagger$ \\
 &
  Other &
  {0.26 $\pm$ 0.18} &
  0.53 $\pm$ 0.14 &
   &
  6.98 $\pm$ 0.77 &
  $\ddagger$ &
  {0.00 $\pm$ 0.00} &
  0.05 $\pm$ 0.11 &
   &
  0.02 $\pm$ 0.05 &
   \\ \hline
\multirow{4}{*}{\rotatebox[origin=c]{90}{Relig. Beli.}} &
  Christian &
  {12.86 $\pm$ 1.35} &
  11.62 $\pm$ 0.65 &
   &
  0.00 $\pm$ 0.00 &
  $\ddagger$ &
  {3.83 $\pm$ 0.90} &
  4.72 $\pm$ 0.86 &
  * &
  0.46 $\pm$ 0.05 &
  $\ddagger$ \\
 &
  Agnostic &
  {84.74 $\pm$ 1.16} &
  84.29 $\pm$ 0.55 &
   &
  94.60 $\pm$ 0.41 &
  $\ddagger$ &
  {91.06 $\pm$ 0.72} &
  90.68 $\pm$ 0.89 &
   &
  93.63 $\pm$ 0.64 &
  $\ddagger$ \\
 &
  Atheist &
  {0.12 $\pm$ 0.12} &
  0.00 $\pm$ 0.00 &
  * &
  4.02 $\pm$ 0.47 &
  $\ddagger$ &
  {3.66 $\pm$ 1.05} &
  2.90 $\pm$ 0.42 &
   &
  5.25 $\pm$ 0.73 &
  $\ddagger$ \\
 &
  Others &
  {2.28 $\pm$ 0.48} &
  4.09 $\pm$ 0.26 &
  $\ddagger$ &
  1.38 $\pm$ 0.27 &
  $\dagger$ &
  {1.45 $\pm$ 0.27} &
  1.69 $\pm$ 0.31 &
   &
  0.66 $\pm$ 0.20 &
  $\ddagger$ \\ \hline
\multirow{3}{*}{\rotatebox[origin=c]{90}{Sex. Or.}} &
  Hetero. &
  {70.07 $\pm$ 1.63} &
  59.57 $\pm$ 1.02 &
  $\ddagger$ &
  5.88 $\pm$ 0.79 &
  $\ddagger$ &
  {99.86 $\pm$ 0.21} &
  99.81 $\pm$ 0.14 &
   &
  99.66 $\pm$ 0.16 &
   \\
 &
  LGBTQ+ &
  {29.93 $\pm$ 1.63} &
  40.41 $\pm$ 1.07 &
  $\ddagger$ &
  94.12 $\pm$ 0.79 &
  $\ddagger$ &
  {0.14 $\pm$ 0.21} &
  0.19 $\pm$ 0.14 &
   &
  0.31 $\pm$ 0.18 &
   \\
 &
  Unspe. &
  {0.00 $\pm$ 0.00} &
  0.02 $\pm$ 0.05 &
   &
  0.00 $\pm$ 0.00 &
   &
  {0.00 $\pm$ 0.00} &
  0.00 $\pm$ 0.00 &
   &
  0.02 $\pm$ 0.05 &
  
\end{tabular}

\end{table*}

\subsection{Analysing sociodemographic attributes generation (RQ1 and RQ2)}

We first ask what sociodemographic characteristics LLMs assume when asked to generate personas from personality-test responses alone (RQ1), and how these assumptions shift under extreme trait manipulation (RQ2). 

Tables \ref{tab:populations_stats_gpt}, \ref{tab:populations_stats_LLaMa}, and \ref{tab:populations_stats_claude} report the mean and standard deviation (calculated across ten trials for the Base populations and five trials each for the MaxP and MaxN populations) across five sociodemographic categories: Gender, Political Orientation, Race, Religious Belief, and Sexual Orientation.\footnote{To analyze model outputs consistently across experiments/trials and models, we normalized the free-text sociodemographic attributes produced by the LLMs into a compact, pre-defined set of categories. More details in 
Appendix \ref{appendix:categories_aggregation}.}
Three additional categories, that is,  age, location, and occupation, were measured, but due to space constraints 
and since differences across populations (Base, MaxN, MaxP) and models were minimal, detailed results are not shown. However, relevant patterns are discussed throughout the paper.

We began by assessing generation consistency, measured as the variance across trials. The low standard deviations observed across categories in Tables \ref{tab:populations_stats_gpt}, \ref{tab:populations_stats_LLaMa}, and \ref{tab:populations_stats_claude} indicate a high degree of consistency for each model.

\begin{table*}[t]
\centering
  \caption{Averages and standard deviations of ten Base, five MaxN and five MaxP sample populations statistics for LLaMa3.2-3B and LLaMa3.1-70B. Cases where the results are significantly different from the Base sample populations are marked as p<0.05*, p <0.01$\dagger$, and p <0.001$\ddagger$ (two-sided t-test). Abbreviations as in Table~\ref{tab:populations_stats_gpt}.}
  \label{tab:populations_stats_LLaMa}
\small  
    \begin{tabular}{@{}l@{\hskip 1pt}|l@{\hskip 1pt}||r@{\hskip 1pt}|r@{}c@{\hskip 1pt}|r@{}c@{}||r@{\hskip 1pt}|r@{}c@{\hskip 1pt}|r@{}c@{}}
& 
   &
  \multicolumn{5}{c||}{LLaMa3.2-3} &
  \multicolumn{5}{c}{LLaMa3.1-70B} \\
  \hline

& 
   &
  \multicolumn{1}{c|}{Base} & 
  \multicolumn{2}{c|}{MaxN} & 
  \multicolumn{2}{c||}{MaxP} &
  \multicolumn{1}{c|}{Base} & 
  \multicolumn{2}{c|}{MaxN} & 
  \multicolumn{2}{c}{MaxP}\\ \hline

\hline
     
\multirow{4}{*}{\rotatebox[origin=c]{90}{Gender}} & 
  Female &
  {0.00 $\pm$ 0.00} &
  0.00 $\pm$ 0.00 &
   &
  0.00 $\pm$ 0.00 &
   &
  {22.03 $\pm$ 2.56} &
  30.07 $\pm$ 0.40 &
$\ddagger$ &
  5.04 $\pm$ 0.90 &
  $\ddagger$ \\
 &
  Male &
  {100.00 $\pm$ 0.00} &
  100.00 $\pm$ 0.00 &
   &
  100.00 $\pm$ 0.00 &
   &
  {77.94 $\pm$ 2.56} &
  69.81 $\pm$ 0.28 &
$\ddagger$ &
  94.34 $\pm$ 0.85 &
  $\ddagger$ \\
 &
  Non-bin. &
  {0.00 $\pm$ 0.00} &
  0.00 $\pm$ 0.00 &
   &
  0.00 $\pm$ 0.00 &
   &
  {0.02 $\pm$ 0.05} &
  0.02 $\pm$ 0.05 &
   &
  0.38 $\pm$ 0.16 &
  $\ddagger$ \\
 &
  Other &
  {0.00 $\pm$ 0.00} &
  0.00 $\pm$ 0.00 &
   &
  0.00 $\pm$ 0.00 &
   &
  {0.00 $\pm$ 0.00} &
  0.10 $\pm$ 0.10 &
  * &
  0.24 $\pm$ 0.23 &
  $\dagger$ \\ \hline
\multirow{4}{*}{\rotatebox[origin=c]{90}{Pol. Or.}} &
  Centre &
  {1.94 $\pm$ 0.33} &
  2.16 $\pm$ 0.68 &
   &
  0.00 $\pm$ 0.00 &
  $\ddagger$ &
  {10.41 $\pm$ 0.64} &
  6.20 $\pm$ 0.61 &
$\ddagger$ &
  0.00 $\pm$ 0.00 &
  $\ddagger$ \\
 &
  Con. &
  {32.56 $\pm$ 1.68} &
  15.35 $\pm$ 1.03 &
$\ddagger$ &
  0.00 $\pm$ 0.00 &
  $\ddagger$ &
  {42.93 $\pm$ 1.23} &
  42.16 $\pm$ 0.75 &
   &
  0.00 $\pm$ 0.00 &
  $\ddagger$ \\
 &
  Prog. &
  {65.37 $\pm$ 1.57} &
  82.42 $\pm$ 1.13 &
$\ddagger$ &
  98.67 $\pm$ 0.52 &
  $\ddagger$ &
  {37.19 $\pm$ 1.69} &
  50.43 $\pm$ 0.67 &
$\ddagger$ &
  69.88 $\pm$ 1.23 &
  $\ddagger$ \\
 &
  Others &
  {0.12 $\pm$ 0.12} &
  0.07 $\pm$ 0.11 &
   &
  1.33 $\pm$ 0.52 &
  $\ddagger$ &
  {9.47 $\pm$ 0.56} &
  1.21 $\pm$ 0.35 &
$\ddagger$ &
  30.12 $\pm$ 1.23 &
  $\ddagger$ \\ \hline
\multirow{5}{*}{\rotatebox[origin=c]{90}{Race}} &
  Asian &
  {0.00 $\pm$ 0.00} &
  0.00 $\pm$ 0.00 &
   &
  0.00 $\pm$ 0.00 &
   &
  {0.90 $\pm$ 0.80} &
  0.36 $\pm$ 0.19 &
  $\dagger$ &
  0.00 $\pm$ 0.00 &
  $\ddagger$ \\
 &
  Black &
  {0.00 $\pm$ 0.00} &
  0.00 $\pm$ 0.00 &
   &
  0.00 $\pm$ 0.00 &
   &
  {0.00 $\pm$ 0.00} &
  0.00 $\pm$ 0.00 &
   &
  0.00 $\pm$ 0.00 &
   \\
 &
  Latin &
  {0.00 $\pm$ 0.00} &
  0.00 $\pm$ 0.00 &
   &
  0.00 $\pm$ 0.00 &
   &
  {0.00 $\pm$ 0.00} &
  0.02 $\pm$ 0.05 &
   &
  0.00 $\pm$ 0.00 &
   \\
 &
  White &
  {100.00 $\pm$ 0.00} &
  100.00 $\pm$ 0.00 &
   &
  100.00 $\pm$ 0.00 &
   &
  {99.08 $\pm$ 0.79} &
  99.52 $\pm$ 0.09 &
  * &
  99.39 $\pm$ 0.31 &
   \\
 &
  Other &
  {0.00 $\pm$ 0.00} &
  0.00 $\pm$ 0.00 &
   &
  0.00 $\pm$ 0.00 &
   &
  {0.02 $\pm$ 0.05} &
  0.10 $\pm$ 0.10 &
   &
  0.61 $\pm$ 0.31 &
  $\ddagger$ \\ \hline
\multirow{4}{*}{\rotatebox[origin=c]{90}{Relig. Beli.}} &
  Christian &
  {76.27 $\pm$ 0.78} &
  72.64 $\pm$ 1.94 &
$\ddagger$ &
  11.45 $\pm$ 1.63 &
  $\ddagger$ &
  {61.16 $\pm$ 0.45} &
  61.72 $\pm$ 1.35 &
   &
  0.00 $\pm$ 0.00 &
  $\ddagger$ \\
 &
  Agnostic &
  {23.24 $\pm$ 0.82} &
  26.51 $\pm$ 1.79 &
$\ddagger$ &
  67.75 $\pm$ 2.40 &
  $\ddagger$ &
  {36.85 $\pm$ 0.90} &
  37.38 $\pm$ 1.18 &
   &
  20.15 $\pm$ 2.03 &
  $\ddagger$ \\
 &
  Atheist &
  {0.48 $\pm$ 0.19} &
  0.85 $\pm$ 0.27 &
  * &
  20.80 $\pm$ 1.23 &
  $\ddagger$ &
  {1.89 $\pm$ 1.08} &
  0.87 $\pm$ 0.31 &
$\ddagger$ &
  79.85 $\pm$ 2.03 &
  $\ddagger$ \\
 &
  Others &
  {0.00 $\pm$ 0.00} &
  0.00 $\pm$ 0.00 &
   &
  0.00 $\pm$ 0.00 &
   &
  {0.10 $\pm$ 0.10} &
  0.02 $\pm$ 0.05 &
   &
  0.00 $\pm$ 0.00 &
  * \\ \hline
\multirow{3}{*}{\rotatebox[origin=c]{90}{Sex. Or.}} &
  Hetero. &
  {100.00 $\pm$ 0.00} &
  100.00 $\pm$ 0.00 &
   &
  100.00 $\pm$ 0.00 &
   &
  {99.98 $\pm$ 0.05} &
  99.90 $\pm$ 0.10 &
   &
  96.76 $\pm$ 0.52 &
  $\ddagger$ \\
 &
  LGBTQ+ &
  {0.00 $\pm$ 0.00} &
  0.00 $\pm$ 0.00 &
   &
  0.00 $\pm$ 0.00 &
   &
  {0.02 $\pm$ 0.05} &
  0.00 $\pm$ 0.00 &
   &
  3.00 $\pm$ 0.44 &
  $\ddagger$ \\
 &
  Unspe. &
  {0.00 $\pm$ 0.00} &
  0.00 $\pm$ 0.00 &
   &
  0.00 $\pm$ 0.00 &
   &
  {0.00 $\pm$ 0.00} &
  0.10 $\pm$ 0.10 &
  * &
  0.24 $\pm$ 0.23 &
  $\dagger$
  
\end{tabular}

\end{table*}

\paragraph{Analysing the Base synthetic populations (RQ1):} Examining Tables \ref{tab:populations_stats_gpt}, \ref{tab:populations_stats_LLaMa}, and \ref{tab:populations_stats_claude} reveals that the synthetic sample populations generated by the models share several key traits:  most are either male or female, with claude-3.5-s and GPT-3.5 skewing female,  LLaMa3.2-3B and LLaMa3.1-70B models skewing male, and GPT-4o showing the most gender diversity, including 29.27$\%$ (± 1.70) non-binary individuals. Racially, the majority are white except for claude-3.5-s, which includes 35.16$\%$ (± 1.72) Asian. Most are heterosexual, except for GPT-4o, which includes 29.96$\%$ (± 1.52) LGBTQ+. Regarding religious beliefs, all models (but to a lesser extent for the LLaMa family, which skew towards Christians),
tend to be agnostic or outside major faiths like Buddhism 
and Hinduism, with no instances of Islam and Judaism.
Politically, they lean centrist or progressive, though LLaMa3.2-3B and LLaMa3.1-70B models include significant conservative portions, that is 42.68$\%$ (± 1.00) and 33.16$\%$ (± 1.30) respectively. Furthermore, the analysis of age, location, and occupation reveals that the Base population primarily consists of individuals aged on average 28 to 32, based mostly in major U.S. metropolitan areas (with London (UK) as the only exception), such as Chicago, New York City, San Francisco, and Los Angeles. They are typically employed in highly educated fields, including Accounting \& Finance, Tech \& Engineering, Creative \& Design, Research \& Science, and Health \& Social Care.

\begin{table}[t]
\small
\centering
  \caption{Averages and standard deviations of ten Base, five MaxN, and five MaxP sample populations statistics for claude-3.5-s. Cases where the results are significantly different from the Base sample populations are marked as p<0.05*, p <0.01$\dagger$, and p <0.001$\ddagger$ (two-sided t-test). Abbreviations as in Table~\ref{tab:populations_stats_gpt}.}
  \label{tab:populations_stats_claude}
    \begin{tabular}{@{}l@{\hskip 1pt}|l@{\hskip 1pt}||r@{\hskip 1pt}|r@{}c@{\hskip 1pt}|r@{}c@{}}
& 
   &
  \multicolumn{5}{c}{claude-3.5-s}  \\
& 
   &
  \multicolumn{1}{c|}{Base} & 
  \multicolumn{2}{c|}{MaxN} & 
  \multicolumn{2}{c}{MaxP}  \\ \hline

\hline
     
\multirow{4}{*}{\rotatebox[origin=c]{90}{Gender}} & Female & 87.84 $\pm$ 1.09 & 98.31 $\pm$ 0.18 & $\ddagger$ & 0.83 $\pm$ 0.10 & $\ddagger$ \\
 & Male & 8.93 $\pm$ 1.19 & 0.29 $\pm$ 0.11 & $\ddagger$ & 2.52 $\pm$ 0.45 & $\ddagger$ \\
 & Non-bin. & 3.22 $\pm$ 0.18 & 1.40 $\pm$ 0.18 & $\ddagger$ & 96.66 $\pm$ 0.48 & $\ddagger$ \\
 & Other & 0.00 $\pm$ 0.00 & 0.00 $\pm$ 0.00 &  & 0.00 $\pm$ 0.00 &  \\ \hline
\multirow{4}{*}{\rotatebox[origin=c]{90}{Pol. Or.}} & Centre & 90.70 $\pm$ 1.23 & 79.18 $\pm$ 1.51 & $\ddagger$ & 0.00 $\pm$ 0.00 & $\ddagger$ \\
 & Con. & 0.56 $\pm$ 0.21 & 0.00 $\pm$ 0.00 & $\ddagger$ & 0.00 $\pm$ 0.00 & $\ddagger$ \\
 & Prog. & 8.67 $\pm$ 1.07 & 20.82 $\pm$ 1.51 & $\ddagger$ & 75.72 $\pm$ 0.38 & $\ddagger$ \\
 & Others & 0.07 $\pm$ 0.07 & 0.00 $\pm$ 0.00 &  & 24.28 $\pm$ 0.38 & $\ddagger$ \\ \hline
\multirow{5}{*}{\rotatebox[origin=c]{90}{Race}} & Asian & 34.82 $\pm$ 1.43 & 29.15 $\pm$ 1.58 & $\ddagger$ & 11.72 $\pm$ 0.42 & $\ddagger$ \\
 & Black & 0.00 $\pm$ 0.00 & 0.00 $\pm$ 0.00 &  & 0.00 $\pm$ 0.00 &  \\
 & Latin & 0.12 $\pm$ 0.08 & 0.22 $\pm$ 0.16 &  & 0.05 $\pm$ 0.07 &  \\
 & White & 62.03 $\pm$ 1.70 & 68.96 $\pm$ 1.79 & $\ddagger$ & 6.13 $\pm$ 0.43 & $\ddagger$ \\
 & Other & 3.03 $\pm$ 0.72 & 1.67 $\pm$ 0.22 & $\ddagger$ & 82.10 $\pm$ 0.63 & $\ddagger$ \\ \hline
\multirow{4}{*}{\rotatebox[origin=c]{90}{Relig. Beli.}} & Christian & 0.07 $\pm$ 0.11 & 0.05 $\pm$ 0.07 &  & 0.00 $\pm$ 0.00 &  \\
 & Agnostic & 99.90 $\pm$ 0.10 & 99.95 $\pm$ 0.07 &  & 99.81 $\pm$ 0.16 &  \\
 & Atheist & 0.00 $\pm$ 0.00 & 0.00 $\pm$ 0.00 &  & 0.19 $\pm$ 0.16 & $\dagger$ \\
 & Others & 0.02 $\pm$ 0.05 & 0.00 $\pm$ 0.00 &  & 0.00 $\pm$ 0.00 &  \\ \hline
\multirow{3}{*}{\rotatebox[origin=c]{90}{Sex. Or.}} & Hetero. & 96.78 $\pm$ 0.18 & 98.60 $\pm$ 0.18 & $\ddagger$ & 0.92 $\pm$ 0.14 & $\ddagger$ \\
 & LGBTQ+ & 3.22 $\pm$ 0.18 & 1.40 $\pm$ 0.18 & $\ddagger$ & 99.08 $\pm$ 0.14 & $\ddagger$ \\
 & Unspe. & 0.00 $\pm$ 0.00 & 0.00 $\pm$ 0.00 &  & 0.00 $\pm$ 0.00 & 
\end{tabular}

\end{table}

\paragraph{Comparing Base,  MaxN and  MaxP (RQ2):} 

We now analyse the synthetic populations generated when maximising the P or N scale of the answers to the input questionnaires.

When comparing the Base, MaxN, and MaxP synthetic sample populations, significant (based on a two-sided t-test) sociodemographic shifts emerge. In detail, the MaxN sample populations show more moderate shifts compared to MaxP, but still share some common trends, notably a general increase in progressive political alignment and a modest increase in occupations related to creativity or education. Claude-3.5-s slightly increases female representation to 98.31$\%$ (± 0.17), approximately doubles progressive presence, and slightly shifts from Tech-related professions to Education-related ones, while GPT-4o shows a milder gender shift with small increases in female and non-binary identities, and a rise in writing \& publishing related professional occupations. LLaMa3.2-3B and LLaMa3.1-70B both boost progressive alignment (by $\approx$18$\%$ and  $\approx$13$\%$, respectively). GPT-3.5 is the most conservative in MaxN, showing minimal deviation from the Base sample population, making it an outlier in terms of sociodemographic change.

MaxP outputs consistently exhibit a marked shift toward progressive political alignment and increased representation of creative occupations, particularly in creative \& design and writing \& publishing roles. 
Notably, GPT-4o and Claude-3.5-S increase the prevalence of LGBTQ+ and non-binary identities; while this does not imply any real-world association, it raises concerns about stereotype-driven generation and potential pathologizing inferences. LLaMa3.1-70B, in contrast, shifts toward a more male-dominated distribution.
Geographically, MaxP samples across models tend to cluster personas in Western and Southern U.S. cities. GPT-3.5 stands out for reducing creative \& design roles and emphasising events \& community-related professions instead; LLaMa3.2-3B and LLaMa3.1-70B models are distinct in driving strong religious shifts in detriment of Christianity, LLaMa3.2-3B toward agnosticism (similarly, but to a lesser extent, GPT-4o), and LLaMa3.1-70B toward Atheism.

 \begin{table*}[!ht]
  \centering
  \caption{EPQR-A scores per model, sample population, and category. Significant differences from the input scores at the individual level: p<0.05* and p <0.01$\dagger$ (two-sided paired t-test) and at the population level: p<0.05$\mathsection$ and p<0.01$\mathparagraph$ (two-sided unpaired t-test).}
  \label{tab:questionnaires_evals}
    \begin{tabular}{@{}l|@{\hskip 2pt}|l||l@{\hskip 2pt}|l@{\hskip 2pt}|l@{\hskip 2pt}|l@{}}
\multicolumn{1}{l}{} &
\multicolumn{1}{l||}{} &
\multicolumn{4}{c}{EPQR-A} \\ 
\cline{3-6}
 
Model &
\multicolumn{1}{c||}{Pop.} &
\multicolumn{1}{c|}{E} &
\multicolumn{1}{c|}{N} &
\multicolumn{1}{c|}{P} &
\multicolumn{1}{c}{L} \\ \hline \hline
\multirow{3}{*}{GPT-4o} &
Base &
\multicolumn{1}{l|}{2.16$\pm$2.85$\dagger$} &
\multicolumn{1}{l|}{3.06$\pm$2.59} &
\multicolumn{1}{l|}{0.87$\pm$1.04} &
5.92$\pm$0.48$\dagger$ \\
&
MaxN &
\multicolumn{1}{l|}{2.15$\pm$2.86$\dagger$} &
\multicolumn{1}{l|}{5.96$\pm$0.29$\dagger$} &
\multicolumn{1}{l|}{0.91$\pm$1.11$\dagger$} &
5.91$\pm$0.57*  \\
&
MaxP &
\multicolumn{1}{l|}{2.17$\pm$2.87$\dagger$} &
\multicolumn{1}{l|}{3.01$\pm$2.73} &
\multicolumn{1}{l|}{4.75$\pm$1.03$\dagger$} &
5.85$\pm$0.72 \\ \hline
\multirow{3}{*}{GPT-3.5} &
Base &
\multicolumn{1}{l|}{2.52$\pm$2.64$\dagger,\mathsection$} &
\multicolumn{1}{l|}{3.32$\pm$2.36$\dagger,\mathsection$} &
\multicolumn{1}{l|}{0.96$\pm$0.93$\dagger,\mathsection$} &
5.93$\pm$0.41$\dagger$ \\
&
MaxN &
\multicolumn{1}{l|}{2.94$\pm$2.62$\dagger,\mathparagraph$} &
\multicolumn{1}{l|}{4.05$\pm$1.76$\dagger,\mathparagraph$} &
\multicolumn{1}{l|}{1.17$\pm$0.95$\dagger,\mathparagraph$} &
5.94$\pm$0.38$\dagger,\mathsection$ \\
&
MaxP &
\multicolumn{1}{l|}{3.71$\pm$2.32$\dagger,\mathparagraph$} &
\multicolumn{1}{l|}{2.61$\pm$2.30$\dagger,\mathparagraph$} &
\multicolumn{1}{l|}{3.12$\pm$1.18$\dagger,\mathparagraph$} &
5.88$\pm$0.49  \\ \hline
\multirow{3}{*}{claude-3.5-s} &
Base &
\multicolumn{1}{l|}{2.19$\pm$2.85$\dagger$} &
\multicolumn{1}{l|}{3.31$\pm$2.44$\dagger$} &
\multicolumn{1}{l|}{0.61$\pm$0.91$\dagger,\mathparagraph$} &
5.90$\pm$0.60  \\
&
MaxN &
\multicolumn{1}{l|}{2.18$\pm$2.86$\dagger$} &
\multicolumn{1}{l|}{6.00$\pm$0.00$\dagger,\mathparagraph$} &
\multicolumn{1}{l|}{0.72$\pm$0.97$\dagger,\mathparagraph$} &
5.87$\pm$0.61 \\
&
MaxP &
\multicolumn{1}{l|}{2.22$\pm$2.85*} &
\multicolumn{1}{l|}{3.15$\pm$2.68*} &
\multicolumn{1}{l|}{5.51$\pm$0.65$\dagger,\mathparagraph$} &
5.42$\pm$1.42$\dagger,\mathparagraph$  \\ \hline
\multirow{3}{*}{LLaMa3.2-3B} &
Base &
\multicolumn{1}{l|}{2.55$\pm$2.48$\dagger,\mathsection$} &
\multicolumn{1}{l|}{2.42$\pm$2.18$\dagger,\mathparagraph$} &
\multicolumn{1}{l|}{1.38$\pm$1.11$\dagger,\mathparagraph$} &
5.22$\pm$0.44$\dagger,\mathparagraph$  \\
&
MaxN &
\multicolumn{1}{l|}{2.48$\pm$2.55$\dagger$} &
\multicolumn{1}{l|}{3.94$\pm$2.06$\dagger,\mathparagraph$} &
\multicolumn{1}{l|}{1.49$\pm$1.05$\dagger,\mathparagraph$} &
5.04$\pm$0.29$\dagger,\mathparagraph$  \\
&
MaxP &
\multicolumn{1}{l|}{3.06$\pm$2.36$\dagger,\mathparagraph$} &
\multicolumn{1}{l|}{2.72$\pm$2.22$\dagger,\mathparagraph$} &
\multicolumn{1}{l|}{2.75$\pm$0.97$\dagger,\mathparagraph$} &
4.99$\pm$0.31$\dagger,\mathparagraph$ \\ \hline
\multirow{3}{*}{LLaMa3.1-70B} &
Base &
\multicolumn{1}{l|}{2.22$\pm$2.83*} &
\multicolumn{1}{l|}{3.49$\pm$2.33$\dagger,\mathparagraph$} &
\multicolumn{1}{l|}{0.75$\pm$1.04$\dagger,\mathsection$} &
5.94$\pm$0.57$\dagger$  \\
&
MaxN &
\multicolumn{1}{l|}{2.24$\pm$2.85} &
\multicolumn{1}{l|}{5.91$\pm$0.33$\dagger,\mathparagraph$} &
\multicolumn{1}{l|}{0.75$\pm$0.97$\dagger,\mathsection$} &
5.95$\pm$0.53$\dagger,\mathsection$  \\
&
MaxP &
\multicolumn{1}{l|}{2.54$\pm$2.77$\dagger,\mathsection$} &
\multicolumn{1}{l|}{2.99$\pm$2.60*} &
\multicolumn{1}{l|}{4.50$\pm$1.46$\dagger,\mathparagraph$} &
5.90$\pm$0.59  \\ \hline \hline
&

Input &
\multicolumn{1}{l|}{2.26$\pm$2.79} &
\multicolumn{1}{l|}{3.08$\pm$2.42} &
\multicolumn{1}{l|}{0.85$\pm$0.97} &
5.89$\pm$0.54 \\ 
&
  Random &
  \multicolumn{1}{l|}{2.23$\pm$1.20} &
  \multicolumn{1}{l|}{3.01$\pm$1.13} &
  \multicolumn{1}{l|}{0.82$\pm$0.72} &
  \multicolumn{1}{l}{5.89$\pm$0.33} 
\end{tabular}
\end{table*}


\subsection{Analysing the degree to which LLM-generated personas reflect the intended personality traits (RQ3)}

Despite large shifts in inferred demographics under trait manipulation, the personas generally preserve the intended personality profiles when evaluated through the same questionnaire that generated them.
We test this trait fidelity by asking each persona to complete the EPQR-A again and comparing regenerated scores to the input distribution (MAE/RMSE), complemented by internal-consistency checks (Cronbach’s $\alpha$); for the Base, population we further probe whether EPQR-A patterns generalise to a second instrument (BFI) via cross-test correlations.


We start by focusing on the Base sample populations, the top rows for each model in Table \ref{tab:questionnaires_evals} report the scores for the Base sample population across the E, N, P, and L dimensions. Although based on the paired t-tests, the observed differences with the input population (second row from bottom) are statistically significant in many cases, 
they were also modest in practical terms, as reflected by the low MAE and RMSE values presented in Table~\ref{tab:accuracy_per_model_generated}. 
Similarly, in MaxN (second rows) and MaxP (third rows), the dimensions not explicitly altered retain values close to the scores of the input questionnaires (Table \ref{tab:questionnaires_evals}, see also Table 
\ref{tab:accuracy_per_model_borderline} in Appendix \ref{appendix:acc&error} 
for low MAE and RMSE scores). For MaxN populations, models such as claude-3.5-s, GPT-4o, and LLaMa3.1-70B reach or nearly approximate the maximum N score (6), while LLaMa3.2-3B and GPT-3.5 show moderate increases, tumbling around 4.

This effect is more pronounced for the P dimension in MaxP, where scores vary widely, from 5.51 (± 0.65) with claude-3.5-s to 2.75 (± 0.97) with LLaMa3.2-3B, averaging around 4.1 across all models. Although apart from the expected differences in the P and N scales for MaxP and MaxN,  significant differences between the input and the newly generated population's measurements can be observed in many scales and models, those differences are small. This is reflected in the relatively small differences in the corresponding MAE and RMSE errors
(Table \ref{tab:accuracy_per_model_borderline}).

Diving into the resemblance between the persona descriptions and the EPQR-A questionnaire, a qualitative analysis reveals that the models generated descriptions closely mirror some questionnaire items. For example a sentence like, "Alex enjoys connecting with people and rarely stays in the background during social occasions" echoes item 15 ("Do you tend to keep in the background on social occasions?" ), while "They are honest to a fault and hold themselves to a high moral standard, always practicing what they preach" reflects item 24 ("Do you always practice what you preach?'').

To assess how surface-level alignment with the EPQR-A affects questionnaire responses, we had the Base population also complete the BFI (scores are reported 
in Table \ref{tab:BFI_score}). 
This enabled cross-test correlation analysis. Table \ref{tab:correlation} shows strong correlations for E across both \mbox{EPQR-A} and BFI (r > 0.94, p < 0.001), and similarly high correlations for N in claude-3.5-s, LLaMa3.1-70B, and GPT-4o (r > 0.91, p < 0.001). Correlations for other models remained adequate. As expected, C from the BFI negatively correlated with P from EPQR-A for claude-3.5-s, GPT-4o, and LLaMa3.1-70B. Interestingly, only LLaMa3.2-3B showed a significant positive correlation between A and P (r = 0.44, p < 0.001). Finally, BFI's O dimension positively correlated with EPQR-A's E.

\begin{table*}[!t]
  \centering
  \caption{Average $\pm$ standard deviations of BFI scores of the Base and input population sample per model.}
  \label{tab:BFI_score}
    \begin{tabular}{@{}l@{\hskip 2pt}|l || l@{\hskip 2pt}|l@{\hskip 2pt}|l@{\hskip 2pt}|l@{\hskip 2pt}|l@{}}
\multicolumn{1}{l}{} &
\multicolumn{1}{l||}{} &
\multicolumn{5}{c}{BFI} \\ 
\cline{3-7}
\multicolumn{1}{c}{Model} &
\multicolumn{1}{c||}{Pop.} &
\multicolumn{1}{c|}{E} &
\multicolumn{1}{c|}{N} &
\multicolumn{1}{c|}{A} &
\multicolumn{1}{c|}{C} &
\multicolumn{1}{c}{O} \\ \hline \hline

GPT-4o &
Base &
{2.81 $\pm$  1.49} &
{3.06 $\pm$ 0.98} &
{4.56 $\pm$ 0.23} &
{4.73 $\pm$ 0.27} &
{4.02 $\pm$ 0.74} \\
 \hline
GPT-3.5 &
Base &
{2.93 $\pm$ 1.08} &
{2.55 $\pm$ 0.57} &
{4.27 $\pm$ 0.27} &
{4.04 $\pm$ 0.21} &
{3.99 $\pm$ 0.42} \\
 \hline
claude-3.5-s &
Base &
{2.78 $\pm$   1.51} &
{3.41 $\pm$ 1.05} &
{4.13 $\pm$ 0.29} &
{4.75 $\pm$ 0.30} &
{3.51 $\pm$ 0.58} \\
\hline
LLaMa3.2-3B &
Base &
{2.93 $\pm$ 0.88} &
{3.03 $\pm$ 0.41} &
{3.94 $\pm$ 0.36} &
{3.96 $\pm$ 0.25} &
{3.32 $\pm$ 0.56} \\
\hline
LLaMa3.1-70B &
Base &
{2.88 $\pm$   1.48} &
{3.10 $\pm$ 0.98} &
{4.45 $\pm$ 0.31} &
{4.70 $\pm$ 0.32} &
{3.19 $\pm$ 0.92} \\
\hline \hline
&
Input 
&
{3.23 $\pm$ 0.71} &
{3.32 $\pm$ 0.42} &
{4.20 $\pm$ 0.43} &
{4.46 $\pm$ 0.37} &
{4.50 $\pm$ 0.31} 
\end{tabular}%

\end{table*}

\begin{table}[!t] 
    \centering
    \caption{Pearson correlation between EPQR-A and BFI, for the Base sample populations per model. Significance \underline{p<0.05} (underline), \textit{p <0.01} (italic), \textbf{p <0.001} (bold).}
        \label{tab:correlation}
\begin{tabular}{@{}l|c||r@{}l|r@{}l|r@{}l|r@{}l|r@{}l@{}}

 & &  \multicolumn{10}{c}{BFI} \\
\cline{3-12}
Model & EPQR-A
   &
  \multicolumn{2}{c|}{E} &
  \multicolumn{2}{c|}{N} &
  \multicolumn{2}{c|}{A} &
  \multicolumn{2}{c|}{C} &
  \multicolumn{2}{c}{O} \\ \hline \hline
\multirow{4}{*}{ GPT-4o}&
  E &
  \textbf{0.99} & &
  \textbf{-0.47}& &
  \textbf{0.45}& &
  \textbf{-0.44}& &
  \textbf{0.26}& \\
\multicolumn{1}{l|}{} &
  N &
  \textbf{-0.38}& &
  \textbf{0.91}& &
  \textbf{-0.33}& &
  \textbf{-0.13}& &
  0.02  \\
\multicolumn{1}{l|}{} &
  P &
  \underline{0.08}&  &
  \textbf{-0.17}& &
  -0.05&&
  \textbf{-0.42}& &
 \textbf{0.71}&\\
\multicolumn{1}{l|}{} &
  L &
  \textit{-0.09}& &
  0.04&&
  \textbf{0.21}& &
  \textbf{0.31}& &
  \textbf{-0.12}& \\ \hline
\multirow{4}{*}{ GPT-3.5}&
  E &
  \textbf{0.96}& &
  \textbf{-0.45}&&
  \textbf{0.29}&&
  0.06& &
  \textbf{0.47}&\\
\multicolumn{1}{l|}{} &
  N &
  \textbf{-0.33}& &
  \textbf{0.75}& &
  \textbf{-0.31}& &
  \textbf{-0.29}& &
  \textbf{-0.18}&\\
\multicolumn{1}{l|}{} &
  P &
  \textbf{0.39}&&
  \textbf{-0.14}&&
  0.06&&
  \textit{-0.09}& &
  \textbf{0.51}&\\
\multicolumn{1}{l|}{} &
  L &
  \textbf{-0.16}& &
  0.04&&
  0.05& &
  \textbf{0.12}&&
  \textit{-0.11}&\\ \hline
\multirow{4}{*}{ claude-3.5-s}&
  E &
  \textbf{0.98}& &
  \textbf{-0.41}& &
  \textbf{0.86}& &
  \textbf{-0.68}& &
  \textbf{0.43}& \\
\multicolumn{1}{l|}{} &
  N &
  \textbf{-0.35}& &
  \textbf{0.93}& &
 \textbf{ -0.37}& &
  \textbf{-0.12}& &
 \textbf{-0.23}& \\
\multicolumn{1}{l|}{} &
  P &
  0.04&  &
  \underline{-0.07}&&
  -0.05& &
  \textbf{-0.18}& &
 \textbf{0.50}&\\
\multicolumn{1}{l|}{} &
  L &
  \textbf{-0.14}&&
  0.03&  &
  0.03&  &
  \textbf{0.25}& &
  \textbf{-0.17}&\\ \hline
\multirow{4}{*}{ LLaMa3.2-3B}&
  E &
  \textbf{0.94}& &
  \textbf{-0.50}&&
  \textbf{0.73}&&
 \textbf{0.12}&&
  \textbf{0.58}&\\
\multicolumn{1}{c|}{} &
  N &
  \textbf{-0.44}&&
  \textbf{0.70}&&
 \textbf{-0.47}&&
 \textbf{-0.33}& &
  \textbf{-0.23}&\\
\multicolumn{1}{l|}{} &
  P &
  \textbf{0.63}&&
  \textbf{-0.39}&&
  \textbf{0.44}&&
  -0.06& &
  \textbf{0.65}&\\
\multicolumn{1}{l|}{} &
  L &
  \textbf{-0.15}&&
  \textbf{-0.15}&&
  -0.01&&
  \textbf{0.16}&&
  \textbf{-0.28}&\\ \hline
\multirow{4}{*}{ LLaMa3.1-70B}&
  E &
  \textbf{0.99}&&
  \textbf{-0.31}&&
  \textbf{0.68}&&
  \textbf{-0.62}&&
  \textbf{0.63}&\\
\multicolumn{1}{l|}{} &
  N &
  \textbf{-0.28}&&
  \textbf{0.93}&&
  \textbf{-0.39}&&
  \underline{-0.08}&&
  \textbf{-0.16}& \\
\multicolumn{1}{l|}{} &
  P &
  \textbf{0.22}& &
  \textbf{-0.15}&&
  -0.04& &
  \textbf{-0.57}&&
  \textbf{0.69}&\\
\multicolumn{1}{l|}{} &
  L &
  \textbf{-0.15}& &
  \underline{0.07}& &
  \textbf{0.20}& &
  \textbf{0.38}& &
  \textit{-0.11}&
\end{tabular}

\end{table}

To assess test consistency, we examined Cronbach’s Alpha. Regarding EPQR-A, Table \ref{tab:cronbach_EPQR-A}, shows that high reliability was observed for E, L, and N (with lower scores in the MaxN samples populations) - except for lama3.2-3B, which showed notably weaker results. In contrast, the P dimension consistently yielded lower reliability, especially in MaxP sample populations. Similarly, 
Table \ref{tab:cronbach_BFI} reports Cronbach’s Alpha score for BFI.

Finally, Table \ref{tab:cronbach_EPQR-A} also reveals that the random baseline lacks internal consistency, with scores around zero. This is expected: although the random baseline mirrors the input sample population's average scores (see Table \ref{tab:questionnaires_evals}), it fails to demonstrate any coherence in personality representation. This contrast suggests that the models analysed here are not just matching trait distributions but are consistently embodying coherent, personality-driven personas.

\begin{table}[!t] 
    \centering
    \caption{Cronbach's Alpha for the EPQR-A test, per model, sample population and category.}
        \label{tab:cronbach_EPQR-A}
\begin{tabular}{l|l||r@{\hskip 2.5pt}|r@{\hskip 2.5pt}|r@{\hskip 2.5pt}|r@{}}
Model        & Pop. & \multicolumn{1}{c|}{E}    & \multicolumn{1}{c|}{N}     & \multicolumn{1}{c|}{P}    & \multicolumn{1}{c}{L}    \\ \hline \hline
\multirow{3}{*}{GPT-4}             & Base  & 1.00& 0.94& 0.61& 0.79\\
                                   & MaxN      & 1.00 & 0.71  & 0.65 & 0.88 \\
                                   & MaxP      & 1.00 & 0.96  & 0.40 & 0.87 \\ \hline
\multirow{3}{*}{GPT-3.5}           & Base  & 0.95 & 0.89  & 0.40 & 0.70 \\
                                   & MaxN      & 0.95 & 0.77  & 0.44 & 0.75 \\
                                   & MaxP      & 0.91 & 0.88  & 0.28 & 0.65 \\ \hline
\multirow{3}{*}{claude-3.5-s} & Base  & 1.00 & 0.92  & 0.59 & 0.86 \\
                                   & MaxN      & 1.00 & -     & 0.60 & 0.80 \\
                                   & MaxP      & 0.99 & 0.95  & 0.23 & 0.89 \\ \hline
\multirow{3}{*}{LLaMa3.2-3B}       & Base  & 0.95 & 0.87  & 0.39 & 0.02 \\
                                   & MaxN      & 0.95 & 0.86  & 0.45 & 0.01 \\
                                   & MaxP      & 0.93 & 0.88  & 0.18 & 0.27 \\ \hline
\multirow{3}{*}{LLaMa3.1-70B}      & Base  & 0.99 & 0.91  & 0.64 & 0.97 \\
                                   & MaxN      & 0.99 & 0.27  & 0.60 & 0.97 \\
                                   & MaxP      & 0.97 & 0.94  & 0.68 & 0.88 \\ \hline \hline
& Input                          & 0.98 & 0.91  & 
0.57 & 0.74
\\
& Random  & 0.03& -0.15& 0.06&0.02\\
\end{tabular}

\end{table}

\begin{table*}[!ht] 
    \centering
    \caption{Cronbach's Alpha for the BFI test for the input population sample and a Base sample population per model.}
        \label{tab:cronbach_BFI}
    \begin{tabular}{@{}l@{\hskip 1pt}|l||l@{\hskip 2.5pt}|l@{\hskip 2.5pt}|l@{\hskip 2.5pt}|l@{\hskip 2.5pt}|l@{}}
Model         & Pop. & E & N& A & C & O\\ 
\hline
\hline
GPT-4             & Base  & 0.99& 0.98& 0.63& 0.87&0.96\\
\hline
GPT-3.5           & Base  & 0.97& 0.86& 0.72& 0.67&0.87\\
 \hline
claude-3.5-s & Base  & 0.99& 0.98& 0.81& 0.9&0.94\\
\hline
LLaMa3.2-3B       & Base  & 0.93& 0.7& 0.84& 0.63&0.87 \\
\hline
LLaMa3.1-70B      & Base  & 0.99& 0.96& 0.84& 0.9&0.97\\
\hline
\hline
                & Input          & 0.96& 0.9& 0.89& 0.92&0.81
\end{tabular}
\end{table*}

\section{Discussion}

\subsection{Answering RQ1 and RQ2}

Our results show that LLMs-generated sample populations lack representativeness, exhibiting strong WEIRD biases~\cite{crockett2023should}. Specifically, the samples predominantly consist of young individuals from Western, Educated, Industrialized, Rich, and Democratic backgrounds. 

To examine whether personality influences the observed biases, we maximised the Neuroticism (N) and Psychoticism (P) dimensions. The results suggest that altering personality traits leads to corresponding shifts in bias, aligning with established findings in personality research.
 
More specifically, when N is maximised, the proportion of women increases (particularly with claude-3.5-s and LLaMa3.1-70B) in line with prior findings linking higher N scores to female respondents - e.g., \cite{chapman2007gender}. Similarly, the proportion of progressive orientation increased, in accordance with literature displaying as high N related with liberal political orientation \cite{gerber2011big} or left-oriented political orientation \cite{krieger2019big}. The rise in progressive political orientation is further amplified in sample populations where P is heightened. This effect would be consistent with studies linking higher psychoticism scores to liberal ideologies \cite{ludeke2016personality}.

An increase in male representation when P is maximised aligns with findings such as \cite{martin1998gender}, though this trend is evident only in LLaMa3.2-3B. In contrast, GPT-4o and claude-3.5-s raise non-binary representation and dramatically increase LGBTQ+ prevalence (only slightly echoed in LLaMa3.2-3B and GPT-3.5). Although recent NLP evidence indicates that LLMs can reproduce negative stereotypes of sexual and gender minorities beyond binary categories \cite{ostrow-lopez-2025-llms}, to our knowledge, no literature links high P with LGBTQ+ identity and this pattern raises concerns. P is associated with traits like aggression and antisocial behaviour, and its alignment with non-binary identities may reflect harmful biases in some LLMs, potentially pathologising these groups.


High P also increases the proportion of creative roles except for GPT-3.5, in accordance with the literature supporting a positive relationship between P measured by the EPQ questionnaire family and creativity \cite{acar2012psychoticism}. The detriment of Christianity observed in both LLaMa models is consistent with the available literature supporting that indices of religiosity are inversely related to P scores \cite{francis2010personality}. 

\begin{figure}[b!]
\centering
\includegraphics[width=11cm]{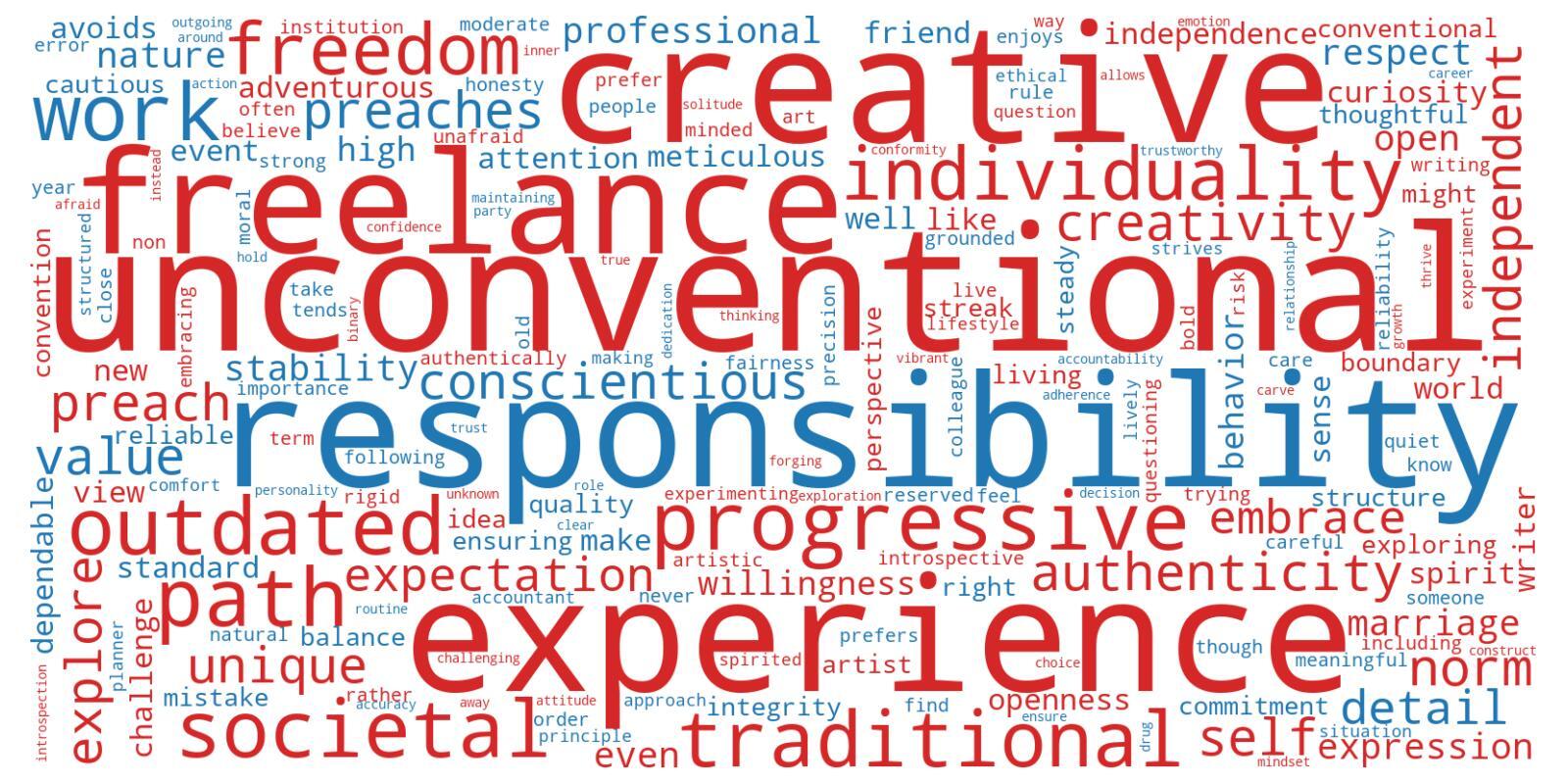}
\caption{Words cloud comparison between description from GPT-4o Base (blue) and from GPT-4o MaxP (red). The size of words is proportional to the absolute difference in frequency: words more frequent in GPT-4o Base compared to  GPT-4o MaxP are colored in blue (red in the opposite case). }
\label{fig:W_cloud}
\end{figure}

It is worth noting that these shifts in demographic distributions appear to reflect the impact of personality traits on the lexical space of persona descriptions. Pairwise word cloud comparisons between Base and MaxN and MaxP sample populations show how personality settings shape word choice (see Figure \ref{fig:W_cloud} for an illustrative example for GPT-4o). While further investigation is needed, this suggests that personality settings may steer the generative space of LLMs.

This interpretation is consistent with findings from persona-steered generation: LLMs can struggle when personas combine multiple traits in ways that are unlikely in human survey data, yielding systematic deviations in the distribution of generated attitudes and attributes \cite{liu-etal-2024-evaluating-large}. 
In our setting, extreme trait anchoring may function similarly by pushing persona descriptions into regions of the model’s generative space where stereotype associations become more salient.

\subsection{Answering RQ3}

Our findings indicate that the models can generate synthetic sample populations that mirror in scores their source personalities. Although the observed differences between the Base, MaxN, and MaxP sample populations and the input population were statistically significant at the individual level (in paired tests), the differences at the population level (unpaired t-tests) are often not significant.
Furthermore, the differences remained relatively small in practical terms, as revealed by an error analysis of MAE and RMSE scores.
Although further investigation is needed, this suggests that while the models effectively replicate source personality traits, they introduce minor deviations, potentially influenced by model guardrails, architecture, and training data.

While all models show strong surface-level similarity between persona descriptions and the \mbox{EPQR-A} items, correlations with BFI responses suggest a deeper embedding of personality traits. High correlations for shared dimensions such as Extraversion (E) and Neuroticism (N) indicate the models effectively capture these traits. A moderate positive correlation between Openness (O) (BFI) and E (EPQR-A) further supports the model’s ability to generalise related traits. However, Psychoticism (P) remains challenging; correlations with Agreeableness (A) were weak or even unexpectedly positive (e.g., in LLaMa3.2-3B), possibly reflecting known reliability issues with the P scale.

Overall, internal consistency is satisfactory for both BFI and EPQR-A questionnaires (except for the P dimension, whose low reliability aligns with known limitations of the EPQR-A instrument in real populations, e.g. \cite{Francis1992, Sandin2002version, Scheibe2023eysenck}). However, when N or  P is maximised, models tend to soften these extremes, resulting in less accurate replication of the intended N and P. Similarly, in the same condition, Cronbach’s alpha values are reduced for N and P, suggesting that exaggerated traits may lead to inconsistencies in persona responses. 

\section{Conclusion}

This paper investigates the viability of generating synthetic sample populations from the scores of personality questionnaires. Our findings reveal both a limitation and a strength of the proposed method. While all models exhibit pronounced WEIRD biases and potentially harmful prejudices against minority groups such as non-binary and LGBTQ+, the results suggest that LLMs can reliably reproduce correlations between personality traits and demographic attributes observed in human sample populations.

The initial observation is especially noteworthy, as our findings indicate that LLMs should be employed with caution when simulating human populations incorporating personality traits, particularly when such populations are used as proxies for real-world diversity.

On the other hand, the second observation is also interesting since our results suggest that LLMs are 
partially
effective at generating synthetic sample populations that reflect the personality traits they were designed to embody (taking into account the models’ difficulties in accurately replicating extreme personality traits, this limitation should be anticipated and addressed in advance when such traits need to be faithfully reproduced).
This result carries particular significance, as grounding personality traits into LLMs may offer a promising path for reducing bias and improving alignment, e.g., \citet{wang2025exploring} shows how altering personality dimensions in LLMs influences output toxicity and bias.

\section{Limitations and Future Work}

This paper presents some limitations that should be subject to future research. Our experiments should be replicated with more languages,  personality tests (e.g., the NEO PI-R \cite{Costa_McCrae_1992}), and other potential variables that can be potentially useful for sample populations generation (e.g., testing the ability of LLMs to generate samples from scores in a questionnaire assessing quality of life). Furthermore, future research should explore patterns arising from other combinations of personality traits, different from MaxN and MaxP, that could be systematically manipulated.

Our method was validated using synthetic sample populations, but future studies should replicate the experiments with human-derived questionnaire data to allow direct comparison between synthetic and real sample populations. 



Finally, another shortcoming of this study, to be addressed in a future research, is the lack of direct comparison with alternative synthetic populations generation methods — such as those based on demographic attributes \cite{argyle2023out, ferreira2024matching, sun2024random}, individual-level data \cite{park2022social}, biographical information of real or fictional figures \cite{shao2023character}, or interviews with real individuals~\cite{park2024generative}.

\section*{Authorship and AI Writing Tools}

For this paper, we used LLM assistance with table formatting and to improve grammar and fluency.




\bibliographystyle{ACM-Reference-Format}
\bibliography{custom}

\appendix

\setcounter{table}{0}
\renewcommand{\thetable}{A\arabic{table}}

\newpage
\section*{APPENDIX}
\section{Example of persona description}\label{appendix:persona_description}
This section presents examples of synthetic personas, including their EPQR-A scores and sociodemographic attributes. Table \ref{tab:exmaple_description_gpt4o} features three GPT-4o personas—one each from the Base, MaxN, and MaxP samples. Table \ref{tab:exmaple_description_base} shows four personas from the Base sample populations generated by GPT-3.5, LLaMa3.2-
3B, LLaMa3.1-70B and claude-3.5-s. The examples from the Base sample populations were generated using the answers to the EPQR-A in Table \ref{tab:exmaple_question_answer}.
The MaxN and MaxP examples were generated from the same answers, with dimensions N and P maximised respectively, as detailed in Section \ref{subsec:experimental_setups}.

\begin{table*}[!ht]

\caption{Example of questions and answers from the EPQR-A questionnaire.} 
 \label{tab:exmaple_question_answer}  
\begin{tabular}{@{}r|p{12.8cm}|r@{}} 

Nr. & Question & Answer  \\
\hline \hline
1 & Does your mood often go up and down? & FALSE
\\
\hline
2 & Are you a talkative person? & FALSE
\\
\hline
3 & Would being in debt worry you? & TRUE
\\
\hline
4 & Are you rather lively? & FALSE
\\
\hline
5 & Were you ever greedy by helping yourself to more than your share of anything?& FALSE
\\
\hline
6 & Would you take drugs which may have strange or dangerous effects? & FALSE
\\
\hline
7 & Have you ever blamed someone for doing something you knew was really your fault? & FALSE
\\
\hline
8 & Do you prefer to go your own way rather than act by the rules? & TRUE
\\
\hline
9 & Do you often feel 'fed-up'? & FALSE
\\
\hline
10 & Have you ever taken anything (even a pin or button) that belonged to someone else? & FALSE
\\
\hline
11 & Would you call yourself a nervous person? & FALSE
\\
\hline
12 & Do you think marriage is old-fashioned and should be done away with? & FALSE
\\
\hline
13 & Can you easily get some life into a rather dull party? & FALSE
\\
\hline
14 & Are you a worrier? & TRUE
\\
\hline
15 & Do you tend to keep in the background on social occasions? & TRUE
\\
\hline
16 & Does it worry you if you know there are mistakes in your work? & TRUE
\\
\hline
17 & Have you ever cheated at a game? & FALSE
\\
\hline
18 & Do you suffer from 'nerves'? & FALSE
\\
\hline
19 & Have you ever taken advantage of someone? & FALSE
\\
\hline
20 & Are you mostly quiet when you are with other people? & TRUE
\\
\hline
21 & Do you often feel lonely? & FALSE
\\
\hline
22 & Is it better to follow society's rules than go your own way? & FALSE
\\
\hline
23 & Do other people think of you as being very lively? & FALSE
\\
\hline
24 & Do you always practice what you preach? & TRUE
\\
\hline \hline 
& \multicolumn{2}{r}{Final Scores: \textbf{E =} 0, \textbf{N =} 1, \textbf{P =} 2, \textbf{L =} 6 }
\end{tabular}
\end{table*}

\begin{table*}[!ht]
\small 

\caption{An example of a persona description, relative EPQR-A scores and sociodemographic attributes generated with GPT-4o for the Neat, MaxP and MaxN sample populations based on the answers of Table~\ref{tab:exmaple_question_answer} .} 
 \label{tab:exmaple_description_gpt4o}  
\begin{tabular}{@{}p{0.1cm}|p{15.5cm}@{}} 

 & Description  \\
\hline

\multirow{22}{*}{\rotatebox[origin=c]{90}{Neat}}& 
Alex Morgan is a thoughtful and introspective individual who values authenticity and independence. Coming from a quiet and reserved background, she has always been more inclined to observe rather than actively participate in highly social or energetic environments. Preferring the peace of working independently, Alex pursued a career as a freelance writer, where she can deeply explore her creative passions without the constraints of an office setting or rigid rules. Though not particularly outgoing, she is someone who consistently practices what she believes in, valuing honesty, integrity, and diligence in her professional and personal life. She takes pride in delivering work free of errors and holds herself to high standards, often worrying when those standards might be compromised. However, she has learned to channel this self-discipline into producing high-quality results rather than dwelling on excessive self-criticism. 

Alex's inclination to follow her own path rather than adhering to societal norms reflects her independent thought and inner strength. She is not one to break rules or engage in unethical behavior, showing a strong moral compass that guides her interactions with others. In her personal life, Alex maintains close, meaningful connections, even if she does not crave constant companionship or large social gatherings. Her independent nature helps her enjoy solitude without feeling lonely, allowing her to find fulfillment within herself. She prefers quiet, reflective moments over lively and bustling environments. Despite this, Alex is not detached—she values other people's ideas and enjoys meaningful one-on-one conversations, seeking depth over surface-level chatter.

Though not particularly lively or exuberant, Alex possesses a steady and grounded demeanor that provides comfort and reliability to those who know her. She has a practical approach to challenges, often choosing responsibility over impulsiveness. While she may worry about small missteps in her work or life, she handles these anxieties with a sense of balance, rarely letting them overpower her ability to act. Alex’s ability to adhere to her personal values, coupled with her introspective and self-disciplined nature, makes her a dependable and creative individual in both her personal and professional pursuits. \\
\cline{2-2} &
\textbf{Age:} 34, \textbf{Gender:} Female, \textbf{Sexual Orientation:} Heterosexual, \textbf{Race:} White, \textbf{Religious Belief:} Agnostic,  \textbf{Occupation:} Writing \& Publishing, \textbf{Political Orientation:} Centre, \textbf{Location:} Boston (MA), \textbf{E =} 0, \textbf{N =} 4,  \textbf{P =} 2, \textbf{L =} 6. \\
\hline
\hline

\multirow{17}{*}{\rotatebox[origin=c]{90}{MaxN}}& 
Alex is a reserved and introspective individual with a strong sense of personal integrity and a preference for self-reliance. As a freelance writer based in Portland, Oregon, he often enjoys working independently and finds fulfillment in pursuing meaningful and creative work. He is not the kind of person to seek the spotlight at social gatherings and tends to stay in the background, preferring quieter settings where he can observe and reflect. Despite his introverted nature, Alex places great value on authenticity and principles, always striving to practice what he preaches and avoiding actions that might compromise his moral beliefs. Throughout his life, he has never cheated, stolen, or taken advantage of others, a testament to his strong ethical foundation. While he values independence and individuality, Alex does occasionally struggle with feelings of loneliness and nervousness. He is a natural worrier, often feeling burdened by mistakes or overwhelmed by uncertainties in life. His nervous tendencies contribute to a heightened sense of caution, and he is firmly against engaging in risky behaviors, such as experimenting with drugs. Financial stability is important to him, and the idea of being in debt is a significant concern. Although he may not be the life of the party or the most talkative person in the room, those close to Alex appreciate his depth, quiet determination, and unwavering honesty. He often feels disillusioned with societal norms and chooses to forge his own path rather than blindly conforming to conventions. While he acknowledges that marriage might feel outdated to some, he respects its place and traditional value. In his independent and reflective life, Alex remains committed to his principles, preferring a thoughtful and introspective approach to navigating the world around him.
\\
\cline{2-2} &
\textbf{Age:} 32 , \textbf{Gender:} Male, \textbf{Sexual Orientation:} Heterosexual, \textbf{Race:} White, \textbf{Religious Belief:} Agnostic,  \textbf{Occupation:} Writing \& Publishing, \textbf{Political Orientation:} Centre, \textbf{Location:} Portland (OR),\textbf{E =} 0, \textbf{N =} 6,  \textbf{P =} 2,  \textbf{L =} 6.\\
\hline
\hline

\multirow{18}{*}{\rotatebox[origin=c]{90}{MaxP}}& 
Alex Carter is a thoughtful and introspective individual, often preferring a quiet and independent life. They are not the type to draw attention in social settings and tend to keep in the background during events. Described by others as reserved but principled, Alex practices what they preach and maintains a strong moral compass, rarely wavering from their beliefs. Although generally calm and steady, they find it more comfortable to carve their own unconventional path rather than sticking to societal norms or traditional rules. For instance, Alex sees marriage as outdated and unnecessary, favoring more modern, flexible expressions of partnership.

Despite their quiet nature, Alex has a bold and curious streak. They are open to experimenting and exploring life’s edges, shown by their willingness to consider experiences like trying substances with potentially strange or dangerous effects. Their liberal political views and artistic approach to life reflect this open-mindedness and desire for freedom. While Alex doesn’t often get caught up in feelings of loneliness or self-doubt, they do tend to worry and spend time carefully reflecting on their decisions and the world around them.

Honesty is a core value for Alex, as they have never cheated, lied to shift blame, or taken advantage of others. Debt doesn’t faze them, and they approach such challenges with a sense of calm and pragmatism. Highly independent, they are unafraid to reject traditional expectations or societal frameworks that don’t align with their values. Living in Portland, Oregon, Alex finds inspiration in the city’s vibrant creative community and its embrace of individuality. As a freelance artist, they thrive on the freedom to chart their own course while contributing meaningful work that reflects their unique perspective and personal integrity.
\\
\cline{2-2} &
\textbf{Age:} 29, \textbf{Gender:} Non-Binary, \textbf{Sexual Orientation:} LGBTQ+, \textbf{Race:} White, \textbf{Religious Belief:} Agnostic,  \textbf{Occupation:} Creative \& Design, \textbf{Political Orientation:} Progressive, \textbf{Location:} Portland (OR), \textbf{Description:}, \textbf{E =} 0, \textbf{N =} 2, \textbf{P =} 6, \textbf{L =} 6. \\

\end{tabular}
\end{table*}

\begin{table*}[!ht]
\small 
\caption{Examples of persona descriptions, relative EPQR-A scores and sociodemographic attributes generated with GPT-3.5, LLaMa3.2-3B, LLaMa3.1-70B and claude-3.5-s for the Base sample populations (using  Table~\ref{tab:exmaple_question_answer}).} 
 \label{tab:exmaple_description_base}  
\begin{tabular}{@{}p{0.1cm}|p{15.6cm}@{}} 

 & Description  \\
\hline

\multirow{13}{*}{\rotatebox[origin=c]{90}{GPT-3.5}}& 
Emily, a 30-year-old Caucasian female, is a Marketing Manager living in an urban area. She identifies as agnostic and holds moderate political views. Emily's personality shines through her responses to various questions. She does not experience frequent mood swings, is not talkative, and is not lively or one to dominate conversations. She worries about being in debt and over mistakes in her work, indicating a responsible and conscientious nature. Emily prefers to go her own way and sticks to her principles, even if it means keeping in the background at social gatherings. Despite not being labeled as lively, she can inject life into dull parties, showcasing her adaptability. While she admits to being a worrier, she also acknowledges that she is mostly quiet in the presence of others, suggesting a reserved yet caring demeanor. Emily values integrity, as she practices what she preaches and does not blame others or cheat to gain an advantage. She believes in following societal rules but also values individuality and independence. Although she expresses occasional feelings of loneliness, she does not let it dictate her actions. Overall, Emily's personality is characterized by responsibility, integrity, adaptability, and a balance between conformity and independent thinking.
\\
\cline{2-2} &
\textbf{Age:} 30, \textbf{Gender:} Female, \textbf{Sexual Orientation:} Heterosexual, \textbf{Race:} White, \textbf{Religious Belief:} Agnostic,  \textbf{Occupation:} Marketing \& Advertising, \textbf{Political Orientation:} Centre, \textbf{Location:} Urban area, \textbf{E =} 1, \textbf{N =} 2,  \textbf{P =} 1, \textbf{L =} 6. \\
\hline
\hline

\multirow{9}{*}{\rotatebox[origin=c]{90}{LLaMa3.2-3B}}& 
John is a reserved individual who values his independence and often prefers to think for himself. He is a hard worker and takes pride in his profession, but can be quite self-critical at times. John is a loyal friend and partner, and he believes in being honest and authentic in his relationships. He is not one for grand gestures or dramatic displays, but he will go to great lengths to support those he cares about. Despite his introverted nature, John is a creative problem-solver and is well-respected in his field for his attention to detail and analytical mind. He is not one to seek the spotlight, but he is always willing to lend a helping hand when needed.
\\
\cline{2-2} &
\textbf{Age:} 32, \textbf{Gender:} Male, \textbf{Sexual Orientation:} Heterosexual, \textbf{Race:} White, \textbf{Religious Belief:} Agnostic,  \textbf{Occupation:} Tech \& Engineering, \textbf{Political Orientation:} Progressive, \textbf{Location:} New York, \textbf{E =} 1, \textbf{N =} 1,  \textbf{P =} 1, \textbf{L =} 5. \\
\hline
\hline

\multirow{11}{*}{\rotatebox[origin=c]{90}{LLaMa3.1-70B}}& 
Alex is a quiet and reserved individual who tends to keep to himself in social situations. He is not one to seek the spotlight or try to be the life of the party. Despite this, he is a hard worker and takes pride in his job as a software engineer. Alex values honesty and integrity, and always tries to practice what he preaches. He is a bit of a worrier and can get anxious about making mistakes in his work, but he is not one to let his nerves get the best of him. Alex is not a fan of following rules just for the sake of following rules, and prefers to think for himself and do what he thinks is right. He is not a risk-taker and would not engage in reckless behavior, such as taking drugs or cheating at games. Alex is a bit of a perfectionist and can get frustrated if he knows there are mistakes in his work. He values his independence and prefers to go his own way rather than follow the crowd. Despite his reserved nature, Alex is a kind and honest person who always tries to do the right thing.
\\
\cline{2-2} &
\textbf{Age:} 30, \textbf{Gender:} Male, \textbf{Sexual Orientation:} Heterosexual ,\textbf{Race:} White, \textbf{Religious Belief:} Agnostic,  \textbf{Occupation:} Tech \& Engineering, \textbf{Political Orientation:} Others, \textbf{Location:} Portland (OR),\textbf{E =} 0, \textbf{N =} 3,  \textbf{P =} 2,  \textbf{L =} 6.\\
\hline
\hline

\multirow{28}{*}{\rotatebox[origin=c]{90}{claude-3.5-s}}& 
Sarah Thompson is a 32-year-old accountant living in Minneapolis, Minnesota. She is known for her quiet demeanor and tendency to keep to herself in social situations. While not particularly talkative or outgoing, Sarah has a strong sense of personal integrity and values honesty in her interactions with others. Growing up in a middle-class family with Irish-American roots, Sarah developed a strong work ethic and a practical approach to life. She takes her responsibilities seriously, often worrying about the quality of her work and the potential for mistakes. This conscientious nature extends to her personal finances as well, with the prospect of being in debt causing her significant concern. Despite her reserved nature, Sarah is not completely averse to social interactions. She simply prefers smaller, more intimate gatherings to large, boisterous parties. When she does attend social events, she tends to stay in the background, observing rather than participating actively. This introverted tendency doesn't mean she's unhappy or lonely; rather, she finds contentment in her own company and a small circle of close friends. Sarah's personality is marked by a streak of independence. She values her ability to think for herself and often prefers to go her own way rather than blindly following rules or societal expectations. However, this independent spirit is balanced by a strong moral compass. She strives to always practice what she preaches and takes pride in her honesty and ethical behavior. In her professional life as an accountant, Sarah's attention to detail and concern for accuracy serve her well. She approaches her work with diligence and care, always striving for perfection. This perfectionist tendency, while sometimes causing her worry, also ensures that she produces high-quality work that she can be proud of. When it comes to her personal life, Sarah holds traditional values in high regard. She believes in the institution of marriage and sees it as a meaningful commitment rather than an outdated concept. Her views on relationships and family life are generally conservative, reflecting her upbringing and personal beliefs. Sarah's cautious nature extends to her approach to risk and new experiences. She is not one to engage in dangerous or unpredictable behaviors, preferring stability and safety. This careful approach to life helps her maintain a sense of control and security, though it may sometimes limit her experiences of spontaneity or adventure. Overall, Sarah is a thoughtful, principled individual who values honesty, hard work, and personal integrity. While she may not be the life of the party, her quiet strength and dependability make her a valued friend and colleague to those who know her well.
\\
\cline{2-2} &
\textbf{Age:} 32, \textbf{Gender:} Female, \textbf{Sexual Orientation:} Heterosexual, \textbf{Race:} White, \textbf{Religious Belief:} Agnostic,  \textbf{Occupation:} Accounting \& Finance, \textbf{Political Orientation:} Centre, \textbf{Location:} Other, \textbf{E =} 0 , \textbf{N =} 3, \textbf{P =} 2, \textbf{L =} 6. \\
\end{tabular}

\end{table*}

\clearpage

\section{Prompt issued to the LLMs}\label{appendix:prompt_issued}

\subsection{Synthetic Persona Generation}
\label{appendix:persona_prompt}

The generation process:
\begin{enumerate}
  \item Takes questionnaire response data from individuals from the input population.
  \item Passes this data to the defined model with specific instructions.
  \item The model is instructed to generate a coherent narrative biography.
  \item Uses the specified expected schema for structured output.
\end{enumerate}

\paragraph{Prompt for all LLMs}
\begin{spverbatim}
Use the provided data, which represents answers given by an individual, to write a coherent and realistic bio in multi-paragraph form.
Do not include or reference the original questions. Describe the individual's background, and life experiences in a continuous narrative style, without bullet points or section headings. 
Reflect how the individual's behaviors, preferences, and reactions align with the data. 
Keep the language clear and concise.
Ensure that the final text provides a unified, flowing description that accurately represents the persona.

Strictly fill each parameter in the JSON structure below with the corresponding information:
{expected_schema}

Output only JSON. Your entire response must be a valid JSON object, with no additional text, commentary, or formatting outside the JSON.
\end{spverbatim}

Where \texttt{expected\_schema} contains:
\begin{spverbatim}
{
  "name": "string",
  "age": "integer",
  "gender": "string",
  "sexual_orientation": "string",
  "race": "string",
  "ethnicity": "string",
  "religious_belief": "string",
  "occupation": "string",
  "political_orientation": "string",
  "location": "string",
  "description": "string"
}
\end{spverbatim}

\subsection{Questionnaire administration prompts}

\subsection*{Description}
After generating personas, they were evaluated using the EPQR-A or the BFI questionnaires. The models:
\begin{enumerate}
  \item take the generated personas as input.
  \item are presented with questionnaires via prompts.
  \item are instructed to respond to questionnaires as if they were the input persona. \footnote{In the prompt below, we present only two questions as an example. The full questionnaires employed in our experiments are available in \citet{Francis1992} for the EPQR-A, and in \citet{john1999big} for the BFI.}
\end{enumerate}


\subsubsection*{EPQR-A questionnaire prompt}
\begin{spverbatim}
You are being asked to complete a questionnaire.

**Questionnaire:** 

{
  "1": "Does your mood often go up and down?",
  "...": "...",
  "24": "Do you always practice what you preach?"
}

**Instructions:**

1. Answer Format: Provide your answers in a single JSON object. Use the question numbers as keys (enclosed in double quotes) and your responses as values.

2. Responses: For each question, answer only with "True" or "False" (as strings, enclosed in double quotes), based on your honest assessment.

3. Order and Keys: Maintain the same order and question numbers as presented in the questionnaire.

4. Explanation: After answering all the questions, add an additional key "explanation" to the JSON object. The value should be a brief explanation (less than 100 words) summarizing the reasoning behind your responses.

5. Output Only JSON: Your entire response should consist solely of the JSON object, without any additional text, markdown formatting, or code blocks. Do not include any commentary outside the JSON.

**Example Response Format:**

{
  "1": "True",
  "2": "False",
  "3": "True",
  "...": "...",
  "24": "False",
  "explanation": "explain your reasoning here"
}
\end{spverbatim}

\subsubsection*{Big Five questionnaire prompt}
\begin{spverbatim}
You are being asked to complete a personality questionnaire.

**Questionnaire:**

{
  "1": "Is talkative",
  "...": "...",
  "44": "Is sophisticated in art, music, or literature"
}

**Instructions:**

For each statement, rate how well it describes you on a scale from 1 to 5:
1 = Disagree strongly  
2 = Disagree a little  
3 = Neither agree nor disagree  
4 = Agree a little  
5 = Agree strongly

- You will receive the questions as a JSON object with numbers as keys and statements as values.
- You must reply exclusively with a JSON object. The JSON should:
    - Use the same question numbers (as string keys) to record your answers.
    - Include an additional key "explanation", containing a brief explanation (less than 100 words) summarizing the reasoning behind your responses.
\end{spverbatim}

\section{Sociodemographic categories aggregation}\label{appendix:categories_aggregation}
To analyze model outputs consistently across experiments/trials and models, we normalized the free-text sociodemographic attributes produced by the LLMs into a compact, pre-defined set of categories.
We treated spelling and formatting variants as equivalent and collapsed them into a canonical form. This step reduces synonym/format variation (e.g., ``nonbinary'', ``non-binary'', ``gender-fluid'') and prevents sparse,  different/singular labels from biasing group statistics. In detail, the categories used in the paper have been aggregated as follows.

\begin{itemize}
\item \textbf{Gender}
    \begin{itemize}
        \item Female = ``female''. 
        \item Male = ``male'', ``man''.
        \item Non-binary = ``genderfluid'', ``gender-fluid'', ``nonbinary'', ``non-binary'', ``gender non-binary'', ``non-conforming''.
        \item Other = ``gender-neutral'', ``neutral'', ``genderqueer'', and any unmapped.
    \end{itemize}
\item \textbf{Political orientation} 
\begin{itemize}
        \item Centre = ``center/centre/centrist/independent/moderate'' variants. 
        \item Conservative = all ``conservative'' variants.
        \item Progressive = Wide synonym set (e.g., ``liberal'', ``left-leaning'', ``moderate-progressive''). 
        \item Other = Non-standard descriptors.
    \end{itemize}
\item  \textbf{Race}
    \begin{itemize}
        \item Asian = ``asian'', ``asian-american''.
        \item Black = ``black'', ``african american'', ``black/african descent''.
        \item Latin = ``hispanic'', ``latino/latina/latinx/latine'', ``hispanic or latino''
        \item White = ``white'', ``caucasian'', ``white/caucasian''.
        \item Other = All other or ambiguous values.
    \end{itemize}
\item  \textbf{Religious belief}
    \begin{itemize}
        \item Christian = ``christian'', ``catholicism''.
        \item Agnostic = ``agnostic''.
        \item Atheist = ``atheist''.
        \item Other = ``islam'', ``buddhist'', ``hinduist''.
    \end{itemize}
\item  \textbf{Sexual orientation}
    \begin{itemize}
        \item Heterosexual = ``heterosexual'', ``straight''.
        \item LGBTQ+ = ``gay'', ``lesbian'', ``bisexual'', ``pansexual'', ``queer'', ``lgbtq+'', ``asexual'', ``demisexual''.
        \item unspecified = ``unknown'', ``unspecified'', ``undisclosed'', ``empty''.
    \end{itemize}
\end{itemize}





\newpage
\section{Accuracy and Error analysis results}\label{appendix:acc&error}

\begin{table*}[!h]

\centering
\caption{Average accuracy and error metrics by EPQR-A scale of a sample Base population with respect to the answers of the input population sample.}
\label{tab:accuracy_per_model_generated}

\begin{tabular}{@{}c|c||cccccc@{}}
Model & Scale & Acc & Precision & Recall & Specificity & \multicolumn{1}{l}{MAE} & RMSE \\ \hline\hline
\multirow{4}{*}{claude-3.5-s} & E & 97.70 & 96.49 & 98.41 & 97.13 & 0.12 & 0.44 \\
                          & N & 94.83 & 91.84 & 98.70 & 90.76 & 0.31 & 0.72 \\
                          & P & 95.68 & 96.82 & 94.59 & 96.81 & 0.25 & 0.65 \\
                          & L & 98.91 & 96.90 & 96.67 & 99.37 & 0.07 & 0.29 \\
 \hline
\multirow{4}{*}{LLaMa3.2-3B} & E & 88.32 & 81.36 & 95.70 & 82.39 & 0.57 & 0.92 \\
                         & N & 75.93 & 83.74 & 65.85 & 86.54 & 1.29 & 1.92 \\
                          & P & 78.63 & 95.50 & 60.74 & 97.05 & 0.87 & 1.18 \\
                          & L & 86.74 & 95.07 & 22.98 & 99.76 & 0.76 & 0.94 \\
 \hline
\multirow{4}{*}{GPT-3.5} & E & 91.40 & 85.78 & 96.74 & 87.12 & 0.47 & 0.94 \\
                         & N & 81.48 & 79.57 & 85.96 & 76.76 & 1.01 & 1.66 \\
                          & P & 89.79 & 87.74 & 92.84 & 86.65 & 0.54 & 0.87 \\
                          & L & 98.26 & 95.86 & 93.81 & 99.17 & 0.09 & 0.36 \\
 \hline
\multirow{4}{*}{GPT-4o} & E & 97.68 & 96.74 & 98.10 & 97.34 & 0.13 & 0.47 \\
                          & N & 93.04 & 93.40 & 93.00 & 93.08 & 0.40 & 0.83 \\
                          & P & 98.20 & 98.25 & 98.21 & 98.20 & 0.10 & 0.35 \\
                          & L & 99.23 & 97.62 & 97.86 & 99.51 & 0.05 & 0.24 \\
 \hline
\multirow{4}{*}{LLaMa3.1-70B} & E & 97.38 & 94.90 & 99.46 & 95.71 & 0.15 & 0.50 \\
                          & N & 88.03 & 83.83 & 95.00 & 80.70 & 0.61 & 1.10 \\
                          & P & 96.63 & 97.76 & 95.54 & 97.75 & 0.20 & 0.56 \\
                          & L & 98.57 & 94.55 & 97.14 & 98.86 & 0.09 & 0.38 \\
\end{tabular}

\end{table*}

\begin{table*}[!ht]
\centering
\caption{
Average accuracy and error metrics by EPQR-A scale of a sample Base population with respect to the answers of the input population sample with maximized N or P Scale.
}
\label{tab:accuracy_per_model_borderline}
\begin{tabular}{@{}c|c|c||ccrrcc@{}}
Model                                     & Population           & Scale & Acc   & Precision & Recall & Specificity & \multicolumn{1}{l}{MAE} & RMSE \\ \hline\hline
\multirow{8}{*}{claude-3.5-s} & \multirow{4}{*}{MaxN} & E     & 97.48 & 96.10     & 98.32  & 96.80       & 0.14                    & 0.50                     \\
                   &                      & N & 51.29 & 51.29 & 100.00 & 0.00  & 2.92 & 3.79 \\
                   &                      & P & 97.58 & 98.42 & 96.78  & 98.40 & 0.15 & 0.48 \\
                   &                      & L & 98.26 & 96.54 & 93.10  & 99.32 & 0.10 & 0.34 \\ \cline{2-9} 
                   & \multirow{4}{*}{MaxP} & E & 97.74 & 96.00 & 99.05  & 96.69 & 0.12 & 0.43 \\
                   &                      & N & 91.51 & 90.80 & 92.84  & 90.10 & 0.50 & 1.01 \\
                   &                      & P & 22.32 & 26.77 & 30.63  & 13.76 & 4.66 & 4.79 \\
                   &                      & L & 91.77 & 71.73 & 84.88  & 93.17 & 0.49 & 1.35 \\ \hline
\multirow{8}{*}{LLaMa3.2-3B}         & \multirow{4}{*}{MaxN}      & E     & 90.19 & 83.60     & 97.01  & 84.72       & 0.50                    & 0.82                     \\
                   &                      & N & 59.40 & 58.13 & 74.51  & 43.50 & 2.31 & 2.99 \\
                   &                      & P & 78.01 & 98.57 & 57.48  & 99.14 & 0.97 & 1.28 \\
                   &                      & L & 83.74 & 72.97 & 6.43   & 99.51 & 0.94 & 1.04 \\ \cline{2-9} 
                   & \multirow{4}{*}{MaxP}     & E & 83.58 & 75.24 & 94.06  & 75.15 & 0.92 & 1.39 \\
                   &                      & N & 71.23 & 74.82 & 66.17  & 76.55 & 1.43 & 2.06 \\
                   &                      & P & 63.08 & 76.27 & 39.50  & 87.35 & 1.96 & 2.29 \\
                   &                      & L & 82.97 & 46.77 & 3.45   & 99.20 & 0.98 & 1.08 \\ \hline
\multirow{8}{*}{GPT-3.5}  & \multirow{4}{*}{MaxN}          & E & 86.68 & 79.88 & 93.70  & 81.05 & 0.77 & 1.46 \\
                   &                      & N & 50.50 & 51.33 & 67.55  & 32.56 & 2.58 & 3.19 \\
                   &                      & P & 89.33 & 86.99 & 92.84  & 85.71 & 0.58 & 0.98 \\
                   &                      & L & 98.26 & 94.56 & 95.24  & 98.88 & 0.10 & 0.38 \\ \cline{2-9} 
                   & \multirow{4}{*}{MaxP}          & E & 74.03 & 65.31 & 88.90  & 62.10 & 1.53 & 2.39 \\
                   &                      & N & 75.04 & 80.31 & 68.02  & 82.44 & 1.35 & 1.94 \\
                   &                      & P & 54.70 & 54.13 & 70.09  & 38.86 & 2.32 & 2.69 \\
                   &                      & L & 97.56 & 93.26 & 92.26  & 98.64 & 0.14 & 0.50 \\ \hline
\multirow{8}{*}{GPT-4o}  & \multirow{4}{*}{MaxN}          & E & 97.42 & 96.43 & 97.83  & 97.09 & 0.15 & 0.51 \\
                   &                      & N & 51.43 & 51.37 & 99.57  & 0.75  & 2.91 & 3.77 \\
                   &                      & P & 97.34 & 96.71 & 98.09  & 96.56 & 0.16 & 0.44 \\
                   &                      & L & 98.81 & 96.21 & 96.79  & 99.22 & 0.07 & 0.34 \\ \cline{2-9} 
                   & \multirow{4}{*}{MaxP}          & E & 97.52 & 96.35 & 98.14  & 97.02 & 0.14 & 0.50 \\
                   &                      & N & 86.20 & 87.40 & 85.41  & 87.03 & 0.75 & 1.25 \\
                   &                      & P & 35.01 & 38.30 & 46.02  & 23.67 & 3.90 & 4.11 \\
                   &                      & L & 97.54 & 90.80 & 95.12  & 98.03 & 0.15 & 0.58 \\ \hline
\multirow{8}{*}{LLaMa3.1-70B}       &\multirow{4}{*}{MaxN}     & E     & 97.88 & 95.58     & 99.86  & 96.29       & 0.12                    & 0.44                     \\
                   &                      & N & 51.67 & 51.51 & 98.86  & 1.99  & 2.88 & 3.72 \\
                   &                      & P & 97.76 & 98.58 & 96.98  & 98.57 & 0.13 & 0.44 \\
                   &                      & L & 98.67 & 95.10 & 97.14  & 98.98 & 0.08 & 0.33 \\ \cline{2-9} 
                   & \multirow{4}{*}{MaxP}     & E & 93.10 & 86.67 & 99.86  & 87.67 & 0.35 & 0.92 \\
                   &                      & N & 87.33 & 88.74 & 86.23  & 88.48 & 0.58 & 0.98 \\
                   &                      & P & 39.16 & 40.05 & 40.10  & 38.21 & 3.65 & 4.02 \\
                   &                      & L & 98.41 & 94.61 & 96.07  & 98.88 & 0.09 & 0.40
\end{tabular}
\end{table*}

\end{document}